\RequirePackage[british]{babel}
\pdfoutput=1
\documentclass[reqno,a4paper,12pt]{amsart}
\usepackage[a4paper,hmargin=2cm,tmargin=3cm,bmargin=3cm]{geometry}
\usepackage[utf8x]{inputenc}
\usepackage{amsmath,amssymb,amstext,amsthm,amscd,mathrsfs,eucal,bm}
\usepackage{graphicx,color}
\usepackage{array}
\usepackage{cite}
\usepackage{hyperref}
\hypersetup{%
  pdftitle   = {Eulerian digraphs and toric Calabi-Yau varieties},
  pdfkeywords = {Calabi-Yau, toric variety, moduli space, supersymmetry, gauge theory, quiver, eulerian, graph theory},
  pdfauthor  = {Paul de Medeiros},
  pdfcreator = {\LaTeX\ with package \flqq hyperref\frqq}
}
\PrerenderUnicode{éÉ}
\newcommand{\fW}{\mathfrak{W}}

\newcommand{\fh}{\mathfrak{h}}

\newcommand{\RR}{\mathbb{R}}
\newcommand{\CC}{\mathbb{C}}

\newcommand{\ZZ}{\mathbb{Z}}

\theoremstyle{plain}

\theoremstyle{definition}

\newcommand{\MUNCH}[1]{\relax}

\allowdisplaybreaks
\begin{document}
\title[Eulerian digraphs and toric Calabi-Yau varieties]{Eulerian digraphs and toric Calabi-Yau varieties}
\author[de Medeiros]{Paul de Medeiros}
\address{School of Mathematics and Maxwell Institute for Mathematical Sciences, University of Edinburgh, Scotland, UK}
\email{p.demedeiros@ed.ac.uk}
\date{\today}
\begin{abstract}
We investigate the structure of a simple class of affine toric Calabi-Yau varieties that are defined from quiver representations based on finite eulerian directed graphs (digraphs). The vanishing first Chern class of these varieties just follows from the characterisation of eulerian digraphs as being connected with all vertices balanced. Some structure theory is used to show how any eulerian digraph can be generated by iterating combinations of just a few canonical graph-theoretic moves. We describe the effect of each of these moves on the lattice polytopes which encode the toric Calabi-Yau varieties and illustrate the construction in several examples. We comment on physical applications of the construction in the context of moduli spaces for superconformal gauged linear sigma models.    
\end{abstract}
\maketitle
\tableofcontents

\section{Introduction and motivation}
\label{sec:introduction}

Supersymmetric quantum field theories in four dimensions often undergo renormalisation group flow to interesting non-trivial fixed points in the infrared which exhibit an enhanced superconformal symmetry. Indeed the study of superconformal field theories in general continues to provide a great deal of progress in modern theoretical physics with superconformal symmetry typically allowing exact calculations of many interesting physical quantities like anomalous dimensions of chiral operators. An important structure in any such theory is the chiral ring of supersymmetric ground states. Classically the chiral ring corresponds to the coordinate ring of an algebraic variety whose points correspond to gauge-invariant monomials in the matter fields of the theory which solve the D- and F-term equations. To understand this structure, it is often convenient to begin by looking at the classical moduli space of gauge-inequivalent superconformal vacua in the theory before considering quantum corrections and conducting a more exact analysis of the full phase structure of the theory. 

When realised as the low-energy description for D3-branes in IIB string theory, many details of the strongly coupled superconformal field theory can also be inferred from the geometry of the holographically dual supergravity background. Indeed, for D3-brane configurations probing a toric conical singularity, this represents by far the most exhaustively studied and best understood class of AdS/CFT dualities \cite{Malda,KlebanovWitten,AdSCFTReview}. Near the toric singularity, the transverse space to the D3-branes corresponds to an affine toric Calabi-Yau three-fold and the data for the singularity encodes both the superpotential and the gauge-matter couplings for the dual superconformal field theory in terms of a quiver representation of the gauge symmetry group \cite{Douglas:1996sw,Douglas:1997de,Beasley:1999uz,Feng:2000mi,Feng:2001xr}. For a single D3-brane, the gauge group is abelian and holography identifies a particular branch of the superconformal vacuum moduli space of the field theory with the aforementioned affine toric Calabi-Yau three-fold in the dual geometry. For multiple coincident D3-branes, the gauge group is nonabelian and the moduli space has a much more complicated structure but typically corresponds to the symmetric product of multiple copies of this geometry. Thus it is often more convenient to begin by considering the aforementioned branch in the moduli space of the abelian theory and systematic analyses have been undertaken in \cite{Forcella:2008bb,Forcella:2008ng} in terms of both the forward algorithm developed in \cite{Feng:2000mi,Feng:2001xr} and subsequent techniques involving dimer models and brane tilings \cite{Hanany:2005ve,Franco:2005rj,Feng:2005gw,Franco:2006gc,Franco:2005sm,Hanany:2005ss}. It is precisely the cancellation of gauge anomalies at one-loop in the superconformal field theory that, in the context of the abelian theory, ensures the first Chern class vanishes for the relevant branch of the moduli space that is to be identified with the dual geometry. In this context, the anomaly cancellation condition is that the quiver representation must be based on a digraph with all vertices balanced (i.e. at each vertex, there must be an equal number of incoming and outgoing arrows). Whence, at least for connected quivers describing gauge-matter couplings for indecomposable superconformal field theories, this is tantamount to the associated digraph being eulerian (i.e. it must contain a circuit traversing each arrow exactly once). The correspondence between superconformal field theories and Calabi-Yau geometries here is certainly not one-to-one in the sense that non-anomalous theories based on different quiver representations can realise the same toric Calabi-Yau moduli spaces -- this is the phenomenon of toric duality, corresponding to Seiberg duality in the associated superconformal field theory \cite{Beasley:1999uz,Feng:2000mi,Feng:2001xr,Beasley:2001zp}. 

Superconformal gauged linear sigma models can provide a convenient physical description of affine toric Calabi-Yau varieties in arbitrary dimensions, following the construction in \cite{WittenGLSM}. It is perhaps most natural to think of a gauged linear sigma model as arising from the dimensional reduction of a supersymmetric field theory in four dimensions involving $n$ vector superfields and $e$ chiral matter superfields with integer charges $Q_{ia}$ under an abelian gauge group, where $i=1,...,n$ and $a=1,...,e$. The additional data needed to specify the classical theory in four dimensions consists of a choice of $n$ real numbers $t_i$ (corresponding to the Fayet-Iliopoulos parameters modifying the D-terms for each $U(1)$ factor in the gauge group) and a gauge-invariant holomorphic function $\fW$ of the matter fields $X_a$ (corresponding to the F-term superpotential). Quantum consistency of the reduced theory in two dimensions is less restrictive than in four dimensions and superconformal invariance of the gauged linear sigma model is guaranteed provided the charges obey $\sum_{a=1}^e Q_{ia} =0$ (e.g. it ensures the R-symmetry of the conformal superalgebra in two dimensions is non-anomalous). The Higgs branch of the space of classical supersymmetric vacua is defined by those constant matter fields which solve both the D- and F-term equations. The D-term equations are $\sum_{a=1}^e Q_{ia} |X_a |^2 = t_i$ and solutions of the F-term equations correspond to critical points of $\fW$. Whence the moduli space of gauge-inequivalent solutions of the D-term equations describes a K\"{a}hler quotient of $\CC^e$ whose first Chern class vanishes precisely as a consequence of $\sum_{a=1}^e Q_{ia} =0$. 

Although any choice of integral charges $Q_{ia}$ obeying $\sum_{a=1}^e Q_{ia} =0$ defines a consistent superconformal gauged linear sigma model, a natural class of solutions is obtained by taking the charges to be associated with a quiver representation based on any eulerian digraph with $n$ vertices and $e$ arrows. From this perspective, the simple class of affine toric Calabi-Yau varieties whose structure we shall investigate in this paper can be described physically as Higgs branches of gauge-inequivalent D-term solutions for superconformal gauged linear sigma models with matter field charges encoded by eulerian digraph quiver representations and with all the Fayet-Iliopoulos parameters set to zero. The motivation for this restriction is that it will allow us to take advantage of some structural results for the class of eulerian digraphs in order to understand the geometrical structure of the associated toric Calabi-Yau varieties in more detail. In particular, we will show how to generate arbitrary eulerian digraphs by iterating combinations of elementary graph-theoretic moves and determine the effect of each these moves on the associated toric Calabi-Yau varieties, or rather on the convex polytopes which encode them. 

The Calabi-Yau geometries we will consider can also be thought of as Higgs branches of gauge-inequivalent D-term solutions for the non-anomalous superconformal abelian quiver gauge theories in four dimensions that were discussed in the second paragraph above. Of course, typically the existence of a non-trivial superpotential in that context means that it is only from the full space of gauge-inequivalent solutions to both the D- and F-term equations that one recovers the toric Calabi-Yau three-fold for the dual geometry. Indeed the strategy in the forward algorithm \cite{Feng:2000mi,Feng:2001xr} is to use the constraints imposed on the superpotential by the toric Calabi-Yau dual geometry to recast the associated F-term equations obeyed by its critical points in the form of D-term equations. Thereby one can recover the relevant branch of the toric Calabi-Yau three-fold moduli space in terms of the Higgs branch of gauge-inequivalent D-term solutions in a certain auxiliary gauged linear sigma model. It is \lq auxiliary' in the sense that its matter content will typically differ markedly from that in the original superconformal field theory in four dimensions and the charges need not be associated with a quiver representation. It is worth remarking that this is a highly non-trivial procedure and the constraints on the superpotential are not guaranteed to be compatible with the quiver representation based on any eulerian digraph -- the admissible ones are characterised more naturally in terms of the dimer models and brane tilings of \cite{Hanany:2005ve,Franco:2005rj,Feng:2005gw,Franco:2006gc,Franco:2005sm,Hanany:2005ss}, whose classification has been initiated in \cite{Davey:2009bp}. Furthermore, for any such admissible quiver representation with $n$ vertices, the results of \cite{Forcella:2008bb,Forcella:2008ng} indicate that the so-called \lq master space' of gauge-inequivalent F-term solutions should itself also contain a branch of maximal dimension describing an affine toric Calabi-Yau variety of dimension $n+2$. Again, the auxiliary gauged linear sigma model description of this branch of the master space need not have charges associated with the quiver representation for an eulerian digraph though our construction would also apply to instances where this is the case.   

The organisation of this paper is as follows. We begin in Section~\ref{sec:graphtheory} with a basic review of graph theory, noting some essential concepts and setting up a consistent notation for the rest of the paper. Section~\ref{sec:generatingeuleriandigraphs} describes the technique that will be used to generate an arbitrary finite eulerian digraph from some combination of four particular graph-theoretic moves discussed in Section~\ref{sec:digraphs}. Section~\ref{sec:toricquivers} begins by defining the quiver representation of an abelian lie group in terms of a finite digraph before describing how this data can be used to define a strongly convex rational polyhedral cone encoding an affine toric variety. The refinement of this construction is then described for an eulerian digraph which is used to define a convex rational polytope encoding an affine toric Calabi-Yau variety. Finally, we derive the effect of each of the four moves on the aforementioned polytopes before concluding with several examples. 

\section{Some graph theory}
\label{sec:graphtheory}
This section contains a brief review of a few basic graph theoretical concepts and results that will be useful in our forthcoming analysis. A more comprehensive introduction to this material can be found in any graph theory textbook such as \cite{BonMur,BJGut,Diestel}.
  
\subsection{Graphs}
\label{sec:graphs}
A {\emph{graph}} $G=(V,E)$ consists of a set of {\emph{vertices}} $V$ and a set of {\emph{edges}} $E$. To each edge in $E$ one must also assign a pair of vertices in $V$ that it connects. An edge connecting a vertex to itself is called a {\emph{loop}}. The {\emph{degree}} ${\mathrm{deg}}(v)$ of a vertex $v \in V$ is the number of edges in $E$ that end on $v$ (with each loop attached to $v$ counting twice in ${\mathrm{deg}}(v)$). If $E$ contains $e$ edges then clearly $\sum_{v \in V} {\mathrm{deg}}(v) = 2e$, whence the number of vertices with odd degree is always even (this is the so-called handshaking lemma). A graph is called $k${\emph{-regular}} if ${\mathrm{deg}}(v) = k$ for all $v \in V$. Thus $kn =2e$ for a $k$-regular graph with $n$ vertices and $e$ edges. The {\emph{cycle graph}} $C_n$ with $n$ vertices is defined such that its edges form the sides of an $n$-sided polygon, whence it is $2$-regular. The {\emph{complete graph}} $K_n$ with $n$ vertices is defined such that each pair of distinct vertices in it are connected by a single edge, whence it is $(n-1)$-regular.    

An edge is called {\emph{simple}} if it is the only one which connects the pair of vertices assigned to it. A graph is then said to be simple if it contains only simple edges (if loops are included here then a simple graph must be loopless). A {\emph{walk}} in $G$ consists of a sequence of vertices in $V$ such that each pair of consecutive vertices in the sequence are connected by an edge in $E$. A walk with no repeated vertices is called a {\emph{path}} and a closed path is called a {\emph{cycle}}. A walk with no repeated edges is called a {\emph{trail}} and a closed trail is called a {\emph{circuit}}. A pair of vertices in $V$ are said to be {\emph{connected}} if there exists a path between them in $G$. A graph is connected if all its vertices are connected. A path (cycle) is called {\emph{hamiltonian}} if it contains every vertex in $V$ exactly once and a graph is called hamiltonian if it admits a hamiltonian cycle. A trail (circuit) is called {\emph{eulerian}} if it traverses every edge in $E$ exactly once and a graph is called eulerian if it admits an eulerian circuit. There appears to be no simple characterisation of hamiltonian graphs though a number of sufficiency theorems have been established which all rely on assuming one has more than some critical number of edges in the graph. By contrast, for any graph $G$, the following statements are equivalent:
\begin{itemize}
 \item {\emph{$G$ is eulerian.}}
 \item {\emph{$G$ is connected and contains no vertices of odd degree.}}
 \item {\emph{$G$ is connected and its edge set can be partitioned into subsets which define edge-disjoint cycle graphs on the vertices of $G$.}}
\end{itemize} 
The equivalence of the first two statements is Euler's theorem. A corollary of the third statement is that any eulerian graph with $n$ vertices and $e$ edges can be obtained by identifying precisely $e-n$ appropriately chosen vertices in the cycle graph $C_e$. The statements above imply that a connected $k$-regular graph is eulerian when $k$ is even (e.g. every connected $2$-regular graph is a cycle graph). The construction of an eulerian circuit in a given eulerian graph can easily be accomplished using Fleury's algorithm. Moreover, the BEST theorem (named after de Bruijn, van Aardenne-Ehrenfest, Smith and Tutte) provides an efficient (i.e. computable in polynomial time) algorithm for computing the number of distinct eulerian circuits in a given eulerian graph.  

\begin{figure}[h!]
\includegraphics[scale=1.2]{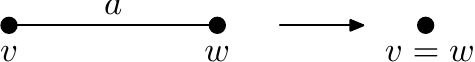} \hspace*{.6in}
\includegraphics[scale=1.2]{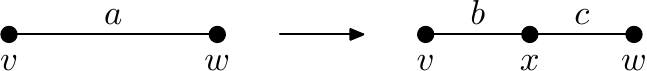}
\caption{Contraction and subdivision of an edge.}
\label{fig1}
\end{figure}

The {\emph{contraction}} of an edge $a \in E$ which connects a pair of distinct vertices $v,w \in V$ in a graph $G$ is defined by first removing $a$ from $E$ (producing the intermediate graph, written $G-a$) and then identifying the vertices $v$ and $w$ to create the new graph, written $G/a$. (Notice that this will create at least one loop at the identified vertex $v=w$ in $G/a$ unless $a$ is simple.) The {\emph{subdivision}} of an edge $a \in E$ connecting vertices $v,w \in V$ in $G$ is defined by adding a new vertex $x$ and two new edges $b$ and $c$ to the graph $G-a$ such that $b$ connects $v$ to $x$ and $c$ connects $x$ to $w$, whence $x$ has degree $2$ in the subdivided graph while the degrees of $v$ and $w$ remain the same as they were in $G$. This can also be thought of as placing an extra vertex on the edge $a$ in between vertices $v$ and $w$. (Notice that the definition of subdivision also applies to loops when $v=w$.) The reverse operation of removing a degree $2$ vertex from a graph (i.e. contracting an edge connected to a degree $2$ vertex) is called {\emph{smoothing}} and we will call a graph {\emph{smooth}} if it contains no vertices with degree $2$. These operations are depicted in Figure~\ref{fig1}.      

Two graphs are defined to be {\emph{homeomorphic}} if they can both be obtained from subdivisions of the same graph (i.e. the operation of subdivision does not modify the topology of a graph). A graph $H$ is a {\emph{subgraph}} of $G$ if its vertex set is a subset of $V$ and its edge set is a subset of $E$ with edges restricted to end on elements in the vertex set of $H$. A graph $H$ is called a {\emph{minor}} of $G$ if it can be obtained by contracting some number of edges in some subgraph of $G$. The notion of graph minor is pivotal in a number of fundamental structure theorems in graph theory. Perhaps the most important being the Robertson-Seymour theorem which states that, in any infinite set of graphs, at least one graph is a (proper) minor of any other. Moreover, any infinite class of graphs that is closed under the operation of taking graph minors can be characterised by a finite set of forbidden minor graphs called the {\emph{obstruction set}} of the class. For example, Wagner's theorem states that the class of planar graphs (i.e. graphs which can be drawn in the plane without any edges crossing) is characterised by an obstruction set consisting of just two elements; the complete graph $K_5$ on five vertices and the complete bipartite graph $K_{3,3}$ on six vertices. Any graph $G$ that is not planar can be drawn without any edges crossing on some closed Riemann surface $\Sigma$ of sufficiently high genus (i.e. the number of unavoidable edge crossings for $G$ drawn in the plane is equal to the minimum number of handles which must be added to the sphere to produce $\Sigma$). The obstruction set for the class of non-planar graphs which can be drawn without edge crossings on the torus is still not known but it must contain more than 16,000 elements! 

\subsection{Digraphs}
\label{sec:digraphs}
An {\emph{orientation}} on a graph is defined by making each edge into an {\emph{arrow}} connecting the same pair of vertices but now with a specified direction pointing from one vertex to the other (a loop based at a vertex in the graph being made into an arrow pointing from the base vertex to itself). A graph $G$ equipped with an orientation is called a {\emph{directed graph}} or {\emph{digraph}}, written $\vec{G}$. For example, a digraph obtained by equipping the complete graph $K_n$ with an orientation is called a {\emph{tournament}}. Many of the concepts and results described above for unoriented graphs can be extended in an obvious way for digraphs. We shall therefore limit our consideration to only those properties which do not translate easily or are refined in a particular way for digraphs. 

One such refinement is that the degree of any vertex $v$ in a digraph can be written ${\mathrm{deg}}(v) = {\mathrm{deg}}^+(v) + {\mathrm{deg}}^-(v)$, where ${\mathrm{deg}}^\pm (v)$ denote the number of arrows pointing from/to $v$ and are respectively referred to as the {\emph{out-/in-degree}} of $v$ (with each loop arrow based at $v$ in the digraph contributing one to both ${\mathrm{deg}}^+(v)$ and ${\mathrm{deg}}^-(v)$). From this definition, it is straightforward to check that $\sum_{v \in V} {\mathrm{deg}}^+(v) = \sum_{v \in V} {\mathrm{deg}}^-(v) = e$ for any digraph with $e$ arrows. A digraph is said to be {\emph{balanced}} if ${\mathrm{deg}}^+(v) = {\mathrm{deg}}^-(v)$ for all $v \in V$. A balanced digraph $\vec{G}$ will be called $k$-regular if ${\mathrm{deg}}^+(v) = k$ for all vertices $v$ (this means that the underlying unoriented graph $G$ is $2k$-regular). Thus $kn =e$ for a $k$-regular balanced digraph with $n$ vertices and $e$ arrows.    

We define an arrow in a digraph $\vec{G}$ to be {\emph{undirected simple}} if the corresponding undirected edge in $G$ is simple. An arrow in $\vec{G}$ pointing from one vertex $v$ to another vertex $w$ is called {\emph{simple}} if it is the only one pointing from $v$ to $w$ (i.e. the arrow is simple even if there is another arrow pointing from $w$ to $v$). Thus any undirected simple arrow must be simple but the converse need not be true. A digraph is then said to be (undirected) simple if it contains only (undirected) simple edges. Thus not every simple digraph is obtained by defining an orientation on a simple graph. A digraph is called {\emph{symmetric}} if, for every arrow pointing from a vertex $v$ to a vertex $w$, there is another arrow pointing from $w$ to $v$. Hence there is a bijective correspondence between symmetric digraphs and graphs, the bijection being the replacement of each pair of oppositely oriented arrows in the symmetric digraph with an unoriented edge in the graph.      

All the different kinds of walks that were defined in a graph $G$ generalise in the obvious way to directed walks in a digraph $\vec{G}$ (i.e. one can only proceed in directions defined by the orientation of the arrows). A digraph $\vec{G}$ is called {\emph{weakly connected}} if its underlying undirected graph $G$ is connected. A digraph $\vec{G}$ is called {\emph{strongly connected}} if it contains a directed path from $v$ to $w$ and a directed path from $w$ to $v$ for all pairs of vertices $(v,w)$. Just as in the undirected case, hamiltonian digraphs are difficult to characterise and eulerian digraphs have a much more straightforward characterisation. That is, for any digraph $\vec{G}$ the following statements are equivalent:
\begin{itemize}
 \item {\emph{$\vec{G}$ is eulerian.}} 
 \item {\emph{$\vec{G}$ is weakly connected and balanced (which implies it is also strongly connected).}} 
 \item {\emph{$\vec{G}$ is strongly connected and its arrow set can be partitioned into arrow-disjoint directed cycles on the vertices of $\vec{G}$.}}  
\end{itemize}
It is perhaps worth emphasising that not every strongly connected digraph is eulerian even though every eulerian digraph is strongly connected. Similar to the undirected case, a corollary of the third statement above is that any eulerian digraph with $n$ vertices and $e$ arrows can be obtained by identifying precisely $e-n$ appropriately chosen vertices in the directed cycle graph $\vec{C}_e$. Note that the cycle digraph is eulerian with the in-degree and out-degree for each of its vertices equal to one (it is the only possible type of $1$-regular eulerian digraph). Clearly the aforementioned identification of vertices in the cycle digraph will produce a new eulerian digraph, with each new identification adding one to both the in- and out-degree of the identified vertex. Both Fleury's algorithm and the BEST theorem can still be used for constructing and counting directed eulerian circuits in eulerian digraphs.        

Evidently the characterisation of eulerian digraphs is a refinement of the characterisation of eulerian graphs due to the additional data provided by the orientation. If $\vec{G}$ is an eulerian digraph then $G$ must be an eulerian graph but clearly the opposite implication follows only provided one chooses an orientation on $G$ such that all its vertices are balanced (i.e. not just of even degree). Moreover, there are typically a number of different balanced orientations that it is possible to equip a given eulerian graph with, leading to a number of distinct eulerian digraphs based on the same underlying eulerian graph. The problem of finding a balanced orientation on an arbitrary eulerian graph, a so-called {\emph{eulerian orientation}}, can be solved efficiently (e.g. for any eulerian circuit on the graph, the order of the traversed edges around the circuit defines an eulerian orientation on the graph). However, the problem of counting the number of distinct eulerian orientations on an arbitrary eulerian graph is difficult (and has been proven to be NP-complete \cite{MikWin1,MikWin2}). Even the problem of counting the number of (edge-labelled) eulerian orientations on a $4$-regular graph is in the same complexity class as the general problem though, in this special case, the number can also be realised in terms of a graph invariant called the Tutte polynomial and as the partition function of a statistical mechanical \lq ice-type' model \cite{Welsh}.

\begin{figure}[h!]
\includegraphics[scale=1.2]{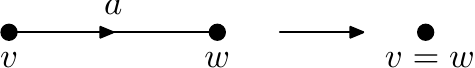} \hspace*{.6in}
\includegraphics[scale=1.2]{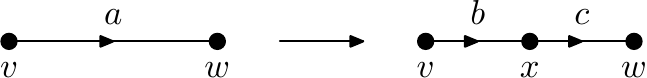}
\caption{Contraction and subdivision of an arrow.}
\label{fig2}
\end{figure}

The contraction and subdivision of an arrow in a digraph are shown in Figure~\ref{fig2} which defines the orientation assignments for these operations relative to Figure~\ref{fig1}.     
Performing either of these operations on any arrow in an eulerian digraph will produce a new digraph that must also be eulerian. The only modification of vertex degrees are such that ${\mathrm{deg}}^+(v) + {\mathrm{deg}}^+(w) -1 = {\mathrm{deg}}^+(v=w)$ for contraction and ${\mathrm{deg}}^+(x)=1$ for subdivision. Only the contraction of an undirected simple arrow will not create a loop in the new digraph while subdivision of an arrow can never create a loop. (Subdivision of a loop based at vertex $v$ creates a $\vec{C}_2$ traversing $v$ and the new vertex $x$.) Contracting an arrow can never create a subdivision. Thus contraction of (unoriented simple) arrows is an operation within the class of (loopless) smooth eulerian digraphs and subdivision of arrows is an operation within the class of loopless eulerian digraphs. It is worth remarking that, if the initial digraph $\vec{G}$ has $n$ vertices and $e$ arrows, then contraction/subdivision of an arrow reduces/increases both of these parameters by one in the new digraph so that their difference is left invariant by both these operations. 

\begin{figure}[h!]
\includegraphics{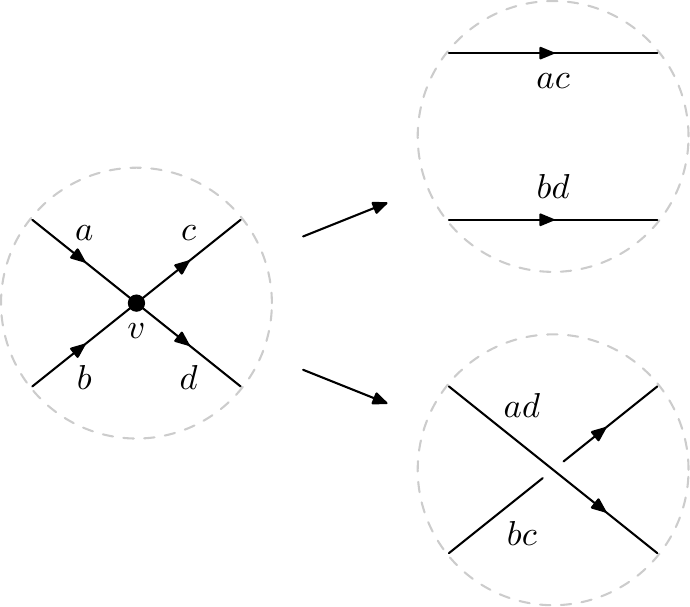}
\caption{Splitting an out-degree $2$ vertex.}
\label{fig3}
\end{figure}

Although the concept of a graph minor can be applied to digraphs, it turns out that this is not such a useful containment relation in the directed case. That is, it is not possible to use this relation to prove a version of the Robertson-Seymour structure theorem for digraphs. In \cite{Johnson}, Johnson proved a structure theorem of this kind for the class of eulerian digraphs using a different containment relation that is more natural in this context and which we will now define. Take any eulerian digraph $\vec{G}$ that contains some positive number of vertices with out-degree $2$. Select one such vertex $v$ in $\vec{G}$ and label its two incoming and outgoing arrows $(a,b)$ and $(c,d)$. The {\emph{splitting}} of vertex $v$ is defined by first deleting $v$ and then connecting the head of $a$ to the tail of either $c$ or $d$ (forming a new arrow $ac$ or $ad$) and the head of $b$ to the tail of either $d$ or $c$ (forming a new arrow $bd$ or $bc$). The two possible outcomes of this operation are shown in Figure~\ref{fig3}.  
The two possible digraphs that can result from splitting $v$ need not be isomorphic but are clearly both eulerian. The choice of which arrows to pair up fixes the eulerian digraph that results from the splitting. The splitting of an out-degree $2$ vertex in a smooth eulerian digraph can never create a subdivision. It is worth noting that, if the initial eulerian digraph has $n$ vertices and $e$ arrows, then splitting reduces $n$ by one and $e$ by two, thus reducing $e-n$ by one. An eulerian digraph $\vec{H}$ is said to be {\emph{immersed}} in an eulerian digraph $\vec{G}$ if $\vec{H}$ can be obtained by applying some number of times to $\vec{G}$ the operations of smoothing, splitting and removing loops. This is the aforementioned containment relation in terms of which Johnson's structure theorem for eulerian digraphs is established.  

\section{Generating eulerian digraphs}
\label{sec:generatingeuleriandigraphs} 
In this section we will show how the graph-theoretic operations that were defined in the previous section can be used to systematically generate any eulerian digraph. The strategy will be described below.

\subsection{Adding loops and subdivisions}
\label{sec:loopandsubdiv} 
Consider first the set of eulerian digraphs $\vec{\mathfrak G}$ and identify any two elements in $\vec{\mathfrak G}$ if they differ only by either the addition of a loop or the subdivision of an arrow (which may be a loop). Within each equivalence class of eulerian digraphs in the set $\vec{\mathfrak F}$ that results from this identification, there is a unique loopless smooth eulerian digraph that can be used to represent the class. The other elements in the class being obtained from all the different possible ways of adding loops and subdividing arrows (including added loops) in the loopless smooth eulerian digraph representative.  Starting from a representative loopless smooth eulerian digraph $\vec{G}$ with $n$ vertices and $e$ arrows, the most general element in the same class that can be obtained from only adding loops is specified by an $n$-tuple of non-negative integers denoting the number of loops to be added at each vertex. Likewise the most general element in the same class that can be obtained only from subdivisions of $\vec{G}$ is specified by an $e$-tuple of non-negative integers denoting the number of subdivisions to be performed on each arrow. Combining and iterating these operations will generate every element in the same class as $\vec{G}$. Of course, depending on the symmetry of $\vec{G}$, different choices of positive integers specifying such operations can give rise to the same eulerian digraph within the class of $\vec{G}$.

\subsection{Loopless smooth eulerian digraphs}
\label{sec:lsedig}
Let us now examine the structure of  $\vec{\mathfrak F}$. By definition, any loopless smooth eulerian digraph $\vec{G} $ representing a class in $\vec{\mathfrak F}$ must contain no subdivisions so that ${\mathrm{deg}}^+(v) >1$ for every vertex $v$ in $\vec{G}$. Consequently the formula $\sum_{v \in V} {\mathrm{deg}}^+(v) = e$ implies that $e \geq 2n$ with $e=2n$ occurring only if $\vec{G}$ is $2$-regular. Let us denote by $\vec{{\mathfrak F}_2} \subset \vec{\mathfrak F}$ the set of all $2$-regular loopless smooth eulerian digraphs and by $\vec{{\mathfrak G}_2} \subset \vec{\mathfrak G}$ the set of all $2$-regular eulerian digraphs. Clearly any element in $\vec{{\mathfrak G}_2}$ must be smooth and there cannot be more than two loops based at any of its vertices. The only possibility of having two loops based at any one vertex is if that is the only vertex in the eulerian digraph. Thus, the only other elements in $\vec{{\mathfrak G}_2}$ that are not in $\vec{{\mathfrak F}_2}$ must contain a single loop on some number of their vertices. Moreover, the removal of any one of these single loops must result in a balanced vertex with out-degree $1$ which is therefore the subdivision of some arrow (which could be a loop). Continuing this procedure of removing a single loop followed by smoothing the resulting subdivision must therefore eventually produce either an element in $\vec{{\mathfrak F}_2}$ or the digraph consisting of two loops based at a single vertex. Conversely, the combined operation of subdividing an arrow (which may be a loop) and adding a loop to the new vertex can therefore be iterated on elements in $\vec{{\mathfrak F}_2}$ and the two loop digraph to obtain any element in $\vec{{\mathfrak G}_2}$. 

\begin{figure}[h!]
\centering
\begin{tabular}{ccccc}
\includegraphics{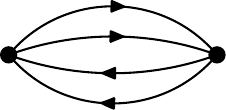} & & & & \\
\includegraphics{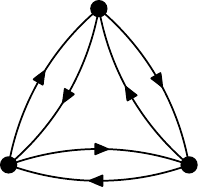} & \includegraphics{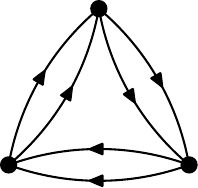}  & & & \\ 
\includegraphics{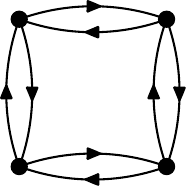} & \includegraphics{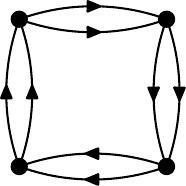}  & \includegraphics{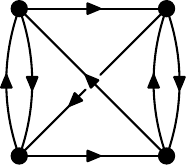} & \includegraphics{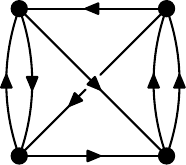} & \includegraphics{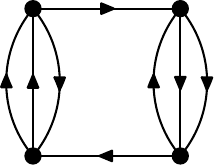} \\ 
\end{tabular}
\caption{Elements in $\vec{\mathfrak F}_2^{[t]}$ are drawn in row $t-1$ for $t=2,3,4$.}  
\label{fig4}
\end{figure}

For each set of eulerian digraphs above, let us define a family of subsets with each member of the family labelled by a superscript $[t]$ and comprising all the elements in the corresponding set which have the same value of $e-n=:t$. As already noted, subdividing or contracting an arrow in a digraph does not change the value of $t$ and so these operations can only map between different elements in the same subset $\vec{\mathfrak G}^{[t]}$. Since any element in $\vec{\mathfrak F}$ has $e \geq 2n$ and $n >1$ then elements in $\vec{\mathfrak F}^{[t]}$ must have $2 \leq n \leq t$ vertices and $e=n+t$ arrows. The elements in $\vec{\mathfrak F}^{[t]}$ with the maximum number of vertices $n=t$ are precisely the $2$-regular ones comprising $\vec{\mathfrak F}_2^{[t]}$. The sets $\vec{\mathfrak F}_2^{[2]}$, $\vec{\mathfrak F}_2^{[3]}$ and $\vec{\mathfrak F}_2^{[4]}$ are shown in Figure~\ref{fig4}. When $t$ is even, there is a single element in $\vec{\mathfrak F}^{[t]}$ with the minimum number of vertices $n=2$ and it is $(\tfrac{t}{2} + 1)$-regular. When $t$ is odd, there are no elements with $n=2$ and $n=3$ is the minimum number of vertices which can be realised for more than one inequivalent eulerian digraph. 

\subsubsection{Contracting undirected simple arrows}
\label{sec:conusa}
Contracting an arrow of some element in $\vec{\mathfrak F}^{[t]}$ will produce an element in $\vec{\mathfrak G}^{[t]}$. This new element is guaranteed to be smooth since contraction can never create a subdivision in a smooth eulerian digraph. However, the new element is also loopless, and hence in $\vec{\mathfrak F}^{[t]}$, only if the the contracted arrow was undirected simple. Thus one can always map from any element in $\vec{\mathfrak F}^{[t]}$ with at least one undirected simple arrow to another element in $\vec{\mathfrak F}^{[t]}$ (with one less vertex and one less arrow) by contracting an undirected simple arrow. 

\begin{figure}[h!]
\includegraphics{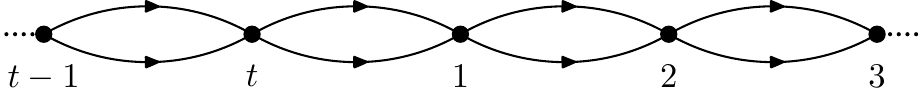} \\ [.2in]
\includegraphics{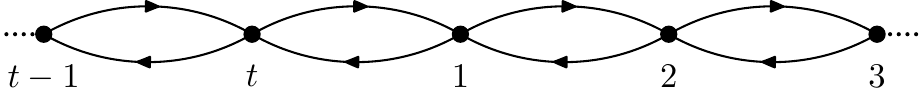}
\caption{The two obstructing elements in $\vec{\mathfrak F}_2^{[t]}$ are \lq necklace' digraphs.}
\label{fig5}
\end{figure}

As we have seen, elements in $\vec{\mathfrak F}_2^{[t]}$ contain the maximum number of vertices and arrows of all the elements in  $\vec{\mathfrak F}^{[t]}$. The only obstruction to being able to obtain by contraction an element in $\vec{\mathfrak F}^{[t]}$ with $t-1$ vertices from an element in $\vec{\mathfrak F}_2^{[t]}$ would be if the latter had no undirected simple arrows. For each value of $t$, there are only two possible elements in  $\vec{\mathfrak F}_2^{[t]}$ that could cause such an obstruction which are depicted in Figure~\ref{fig5}. They can be thought of as the union of two copies of the cycle digraph $\vec{C}_t$ with the same or opposite orientations around the same $t$ vertices (the two orientations being isomorphic only for $t=2$).  
The element in $\vec{\mathfrak F}^{[t]}$ with $t-1$ vertices obtained by contracting an undirected simple arrow in any element in $\vec{\mathfrak F}_2^{[t]}$ (other than one of the two in Figure~\ref{fig5}) will have $t-2$ vertices with out-degree $2$ and one vertex with out-degree $3$ (i.e. the one which formed the endpoints of the contracted arrow). Indeed this must be the case for every element in $\vec{\mathfrak F}^{[t]}$ with $t-1$ vertices. That is, for any element $\vec{G}$ in $\vec{\mathfrak F}^{[t]}$ with vertex set $V$, a more convenient form of the hand-shaking lemma is $\sum_{v \in V} k(v) = t$, where $k(v) := {\mathrm{deg}}^+(v) -1$ is a positive integer for every vertex $v \in V$ as a consequence of $\vec{G}$ being smooth. Thus, when $|V|=t-1$, the only solution is when one vertex has out-degree $3$ with all the rest having out-degree $2$. It is also useful to note that the relation  ${\mathrm{deg}}^+(v) + {\mathrm{deg}}^+(w) -1 = {\mathrm{deg}}^+(v=w)$ between the out-degrees of the vertices $v$ and $w$ which are identified under the contraction of an arrow connecting them can be more conveniently rewritten as $k(v) + k(w) = k(v=w)$. Now, for any element in $\vec{\mathfrak F}^{[t]}$ with $t-1$ vertices, it is possible to write $k(x)=2=1+1$ for the single out-degree $3$ vertex $x$ which means that this element can be obtained by the contraction of an undirected simple arrow in (at least) one element in $\vec{\mathfrak F}_2^{[t]}$. The element in $\vec{\mathfrak F}_2^{[t]}$ is constructed by simply partitioning the six arrow endpoints on $x$ into two sets of three; one set consisting of two outgoing and one incoming arrow endpoints and the other set consisting of one outgoing and two incoming arrow endpoints. Deleting $x$ and having the two sets of endpoints on two new vertices $v$ and $w$ then produces the element in $\vec{\mathfrak F}_2^{[t]}$ after adding a single new arrow $a$ connecting $v$ and $w$ whose orientation is fixed by requiring $v$ and $w$ to be balanced. The element in $\vec{\mathfrak F}^{[t]}$ with $t-1$ vertices is then recovered by contracting $a$. Clearly there are a number of distinct options for partitioning the arrow endpoints in this way leading to a number of different  elements in $\vec{\mathfrak F}_2^{[t]}$ from which the same element in $\vec{\mathfrak F}^{[t]}$ with $t-1$ vertices can be obtained via the contraction of an undirected simple arrow. Nonetheless, the point is to have established that every element in $\vec{\mathfrak F}^{[t]}$ with $t-1$ vertices can be obtained by contracting an undirected simple arrow in some element in $\vec{\mathfrak F}_2^{[t]}$. 

It is straightforward to extend this argument to show that in fact every element in $\vec{\mathfrak F}^{[t]}$ with fewer than $t$ vertices can be obtained from some element in $\vec{\mathfrak F}_2^{[t]}$ (other than one of the two in Figure~\ref{fig5}) by performing some sequence of contractions of unoriented simple arrows. The number of vertices in the resulting element being precisely $t$ minus the number of contractions performed. The trick is to again start with any element in $\vec{\mathfrak F}^{[t]}$ with fewer than $t$ vertices and pick some vertex $x$ with $k(x) >1$. (Such a vertex must exist or else the element would necessarily have $t$ vertices.) Next write $k(x) = p+q$ in terms of positive integers $p$ and $q$ and partition the $2(k(x)+1)$ arrow endpoints on $x$ into two sets; one set consisting of $p+1$ outgoing and $p$ incoming endpoints and the other set consisting of $q$ outgoing and $q+1$ incoming endpoints. Delete $x$ and reattach the two sets of endpoints on two new vertices $v$ and $w$ and connect these two vertices with a single appropriately oriented arrow $a$ such that they are both balanced (the out-degrees of the two new balanced vertices being $p+1$ and $q+1$). This construction therefore produces a new element in $\vec{\mathfrak F}^{[t]}$ with one extra vertex and one extra edge and at least one undirected simple arrow $a$ whose contraction gives back the original element. Again, for a given choice of positive integers $p$ and $q$, there will generally be a number of inequivalent partitions possible for the arrow endpoints leading to a number of different new elements for which the contraction of an undirected simple arrow will give back the same element we started with. This procedure can be iterated for all the different vertices $x$ with $k(x)>1$ of a given element in $\vec{\mathfrak F}^{[t]}$ until eventually one must stop when an element in $\vec{\mathfrak F}_2^{[t]}$ is obtained, thus establishing the result claimed above. Another way of stating this result is to say that every element in $\vec{\mathfrak F}^{[t]}$ is a minor of some element in  $\vec{\mathfrak F}_2^{[t]}$. It is important to stress that this is not a manifestation of the Robertson-Seymour structure theorem for the finite class $\vec{\mathfrak F}^{[t]}$ since not every element in $\vec{\mathfrak F}^{[t]}$ has a (proper) minor that is also contained in $\vec{\mathfrak F}^{[t]}$. Indeed the elements in $\vec{\mathfrak F}^{[t]}$ which do not are precisely those loopless smooth eulerian digraphs with no undirected simple arrows.

\subsubsection{Loopless splitting of $2$-regular vertices}
\label{sec:split2regvert}

The preceding analysis implies that  $\vec{\mathfrak G}$ can be generated by applying to each set $\vec{\mathfrak F}_2^{[t]}$ (and the trivial graph with one vertex and no arrows) all possible combinations of the three operations of loop addition, subdivision and contraction of undirected simple arrows. The fourth and final operation that will be required involves the splitting of a vertex in an element in $\vec{\mathfrak F}_2 = \bigsqcup_{t=2}^\infty \vec{\mathfrak F}_2^{[t]}$.

\begin{figure}[h!]
\centering
\begin{tabular}{cccc}
\includegraphics{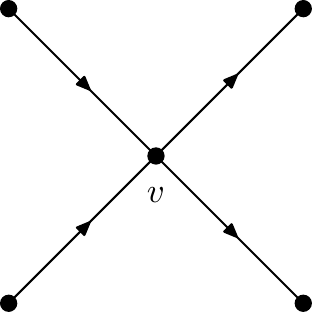} & \includegraphics{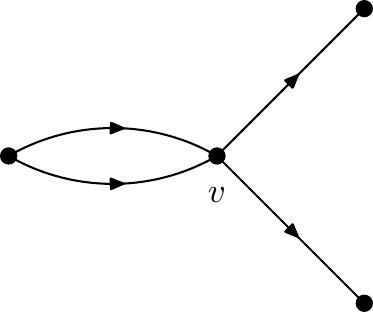} & \includegraphics{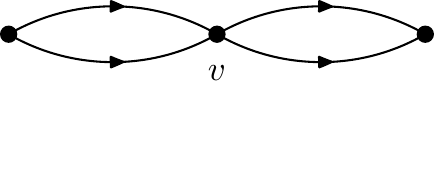} & \\ [.1in]
\includegraphics{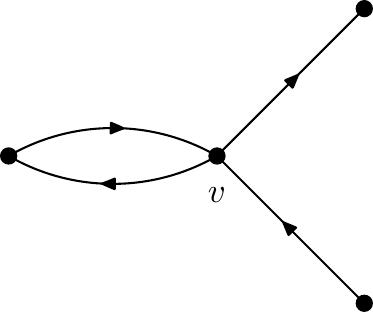} & \includegraphics{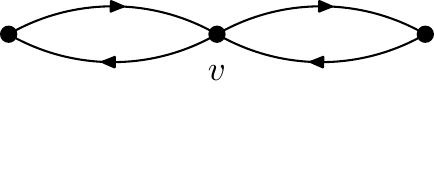} & \includegraphics{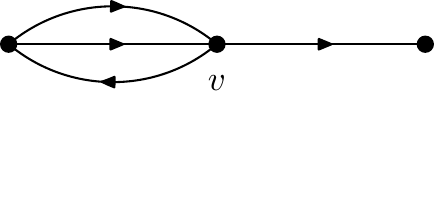} & \includegraphics{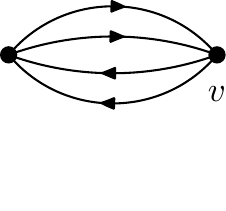} \\ 
\end{tabular}
\caption{The inequivalent arrow connections for a $2$-regular vertex $v$.}  
\label{fig6}
\end{figure}

\begin{figure}[h!]
\includegraphics{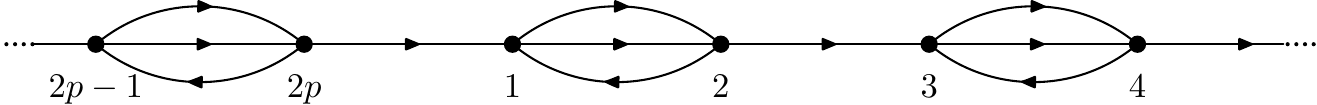}
\caption{The unique obstructing element $\vec{O}_p$ in $\vec{\mathfrak F}_2^{[2p]}$.}
\label{fig7}
\end{figure}

\begin{figure}[h!]
\includegraphics{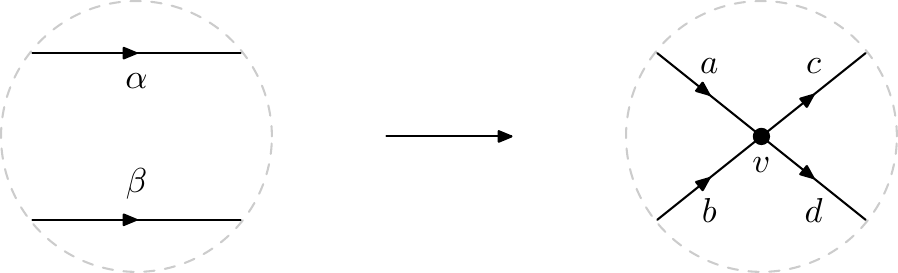}
\caption{Simple immersion of $\vec{H}$ in $\vec{G}$.}
\label{fig8}
\end{figure}

Splitting a vertex in an element in $\vec{\mathfrak F}_2^{[t]}$ will produce an element in $\vec{\mathfrak G}_2^{[t-1]}$. Whichever of the two possible arrow reconnections are chosen in the splitting, the new element is necessarily smooth but need not be loopless. The conditions imposed by requiring the new element to be in $\vec{\mathfrak F}_2^{[t-1]}$ depend on the nature of the connections with other vertices that are made by the four arrows attached to the vertex $v$ to be split. There are seven distinct scenarios that are drawn in Figure~\ref{fig6} (only the four arrows attached to $v$ and the other vertices they are connected to are shown).
Let us define the splitting of a vertex in an element in $\vec{\mathfrak F}_2^{[t]}$ to be loopless if it produces an element in $\vec{\mathfrak F}_2^{[t-1]}$. In the three cases shown in the first row of Figure~\ref{fig6}, either of the two arrow reconnections will give a loopless splitting of $v$.  In the first two cases in the second row, only one arrow reconnection will produce a loopless splitting of $v$. In the last two cases in the second row, no loopless splittings of $v$ are possible. Therefore any element in  $\vec{\mathfrak F}_2^{[t]}$ must contain a vertex that admits a loopless splitting except if all its vertices look like the last two scenarios in Figure~\ref{fig6}. Only for each even value of $t=2p$ is there such an obstructing element $\vec{O}_p$ in $\vec{\mathfrak F}_2^{[2p]}$ which is unique and is depicted in Figure~\ref{fig7} (this includes the unique element in $\vec{\mathfrak F}_2^{[2]}$ when $p=1$). 
Repeating the operation of loopless splitting on as many vertices as possible in any element in $\vec{\mathfrak F}_2^{[t]}$ therefore must eventually produce the digraph $\vec{O}_q$ for some value of $q \leq \lfloor \tfrac{t}{2} \rfloor$. Conversely, this means that any element in $\vec{\mathfrak F}_2$ can be obtained from some $\vec{O}_q$ by reversing the loopless splitting procedure some number of times. That is, by isolating pairs of arrows and joining them to form a new $2$-regular vertex according to the reversal of the diagram in Figure~\ref{fig3} (which will never create any loops or subdivisions).  This reverse operation is shown in Figure~\ref{fig8}  acting on an isolated pair of different arrows $\alpha$ and $\beta$ in some $\vec{H} \in \vec{\mathfrak F}_2^{[t]}$ to produce a new digraph $\vec{G} \in \vec{\mathfrak F}_2^{[t+1]}$ which is the same as $\vec{H}$ everywhere except within the dashed circle. The tails of the new arrows $a$ and $b$ in $\vec{G}$ touch the same vertices as the tails of  $\alpha$ and $\beta$ in $\vec{H}$  while the heads of the new arrows $c$ and $d$ in $\vec{G}$ touch the same vertices as the heads of  $\alpha$ and $\beta$ in $\vec{H}$. 
The reversal of a single loopless vertex splitting in this manner just corresponds to the simplest non-trivial case of an immersion (i.e. it involves no loop additions nor subdivisions here since it is an operation within $\vec{\mathfrak F}_2$).  

The objective of generating all the elements in $\vec{\mathfrak F}_2$ from those in $\{ \vec{O}_p \mid p \in \ZZ_{>0} \}$ by iterating the simple immersion operation above dictated that it need only act on elements in $\vec{\mathfrak F}_2$. However,  in order generate the elements in $\{ \vec{O}_p \mid p \in \ZZ_{>0} \}$ themselves, it is useful to note that one can also define an immersion of $\vec{O}_p$ in $\vec{O}_{p+1}$ in the following way. Start by choosing any undirected simple arrow in $\vec{O}_p$, subdivide it once and then attach a loop to the new vertex. Finally, apply the simple immersion shown in Figure~\ref{fig8} to this intermediate digraph such that $\alpha$ is identified with the aforementioned loop and $\beta$ identified with the arrow whose tail is attached to its base. The resulting digraph is isomorphic to $\vec{O}_{p+1}$. 
 
In summary, we have found that every element in $\vec{\mathfrak G}$ can be generated by applying to the trivial graph some combined iteration of just four moves:\\

\noindent I. Loop addition, mapping $\vec{\mathfrak G}^{[t]}  \rightarrow \vec{\mathfrak G}^{[t+1]} $.

\noindent II. Subdivision of an arrow (see Figure~\ref{fig2}), mapping $\vec{\mathfrak G}^{[t]}  \rightarrow \vec{\mathfrak G}^{[t]} $.

\noindent III. Contraction of an undirected simple arrow (see Figure~\ref{fig2}), mapping $\vec{\mathfrak G}^{[t]}  \rightarrow \vec{\mathfrak G}^{[t]} $.

\noindent IV. Simple immersion (see Figure~\ref{fig8}), mapping $\vec{\mathfrak G}^{[t]}  \rightarrow \vec{\mathfrak G}^{[t+1]} $.\\

By definition, moves I and II generate $\vec{\mathfrak G}$ from  $\vec{\mathfrak F} = \bigsqcup_{t=2}^\infty \vec{\mathfrak F}^{[t]}$. For each value of $t$, move III generates  $\vec{\mathfrak F}^{[t]}$ from $\vec{\mathfrak F}_2^{[t]}$. The set  $\vec{\mathfrak F}_2  = \bigsqcup_{t=2}^\infty \vec{\mathfrak F}_2^{[t]}$ is generated from $\{ \vec{O}_p \mid p \in \ZZ_{>0}  \}$ using move IV and $\{ \vec{O}_p \mid p \in \ZZ_{>0} \}$ is generated using the combination of moves II+I+IV described above. Of course, moves I, II and IV collectively generate the more general operation of immersion for eulerian digraphs. However, it will be convenient in the forthcoming analysis to distinguish the different moves in this way.  

The motivation for describing this graph-theoretic construction will become clear in the next section where we will use eulerian digraphs as the data defining a certain class of affine toric Calabi-Yau varieties. The idea will be to identify precisely what effect these four moves have on the associated toric geometry and thereby say something about the generic structure. It will aid our forthcoming geometrical interpretation of move IV to conclude this section by summarising a neat way of encoding $2$-regular eulerian digraphs.         

\subsection{Encoding $2$-regular eulerian digraphs}
\label{sec:encoding2reg}

As already mentioned in Section~\ref{sec:digraphs}, any eulerian digraph $\vec{G}$ with $n$ vertices and $e$ arrows can be obtained from the cycle digraph $\vec{C}_e$ by identifying $e-n$ appropriately chosen vertices. It worth emphasising that what this identification really produces is a particular eulerian circuit $\vec{\Gamma}$ in $\vec{G}$ (which is of course sufficient to define it). By assigning labels $( i_1 i_2 ... i_e )$ to the vertices cyclically ordered around $\vec{C}_e$, such that each $i_a \in \{ 1,2,...,n \}$ and precisely $e-n$ of the labels are repeated according to the identification above, one specifies a labelling for the vertices in $\vec{\Gamma}$. A labelling for the arrows in $\vec{\Gamma}$ can be specified by simply inserting a different label $m_a \in \{ 1,2,...,e \}$ for each arrow between each successive pair of vertices $i_a$ and $i_{a+1}$ in the cyclically ordered sequence above. We will generally not benefit from distinguishing different labellings of the same eulerian digraph but fixing a labelling can be useful for the purposes of distinguishing different circuits within a given eulerian digraph.

For a $2$-regular eulerian digraph $\vec{G}$, the situation is somewhat simpler. Since $\vec{G}$ has $e=2n$ then each of the $n$ distinct vertex labels must appear exactly twice in the sequence $( i_1 i_2 ... i_{2n} )$. Therefore one can depict any eulerian circuit $\vec{\Gamma}$ in $\vec{G}$ by a circle graph on $2n$ vertices with $n$ chords connecting the $n$ identified vertices within it. We will take the orientation for $\vec{\Gamma}$ to be defined by going clockwise around the circle. In terms of this picture, a pair of different vertices $i$ and $j$ are said to be {\emph{interlaced}} if they appear in the order $(... i ... j ... i ... j ...)$ in $\vec{\Gamma}$ (i.e. if the chords associated with $i$ and $j$ intersect each other). An important insight due to Arratia, Bollob\'{a}s and Sorkin in \cite{InterlaceABS} is that one can utilise the structure of such interlacings as a means of enumerating circuits in $2$-regular eulerian digraphs. The essential ingredient here is called the {\emph{interlace graph}} $I( \vec{\Gamma} )$ of the eulerian circuit $\vec{\Gamma}$ in $\vec{G}$. It is defined as the undirected graph on $n$ vertices such that any pair of vertices $i$ and $j$ are connected by an edge only if they are interlaced in $\vec{\Gamma}$.  The interlace graph is necessarily simple since any pair of vertices can obviously only be interlaced once.

A $2$-regular eulerian digraph is in $\vec{\mathfrak F}_2$ only if it is loopless. This condition just means that the same vertex can never appear consecutively in a sequence of vertices associated with any eulerian circuit. In the notation of Figure~\ref{fig8}, the effect of move IV on an element $\vec{H}$ in $\vec{\mathfrak F}_2$, is easily visualised in terms of the chord diagram obtained from an eulerian circuit $\vec{\Gamma}_{\vec{H}}$ in $\vec{H}$ by replacing  $( ... \alpha ... \beta ...)$ with  $( ... a v c ... b v d ...)$ to form an eulerian circuit $\vec{\Gamma}_{\vec{G}}$ in $\vec{G}$. That is, it corresponds to inserting in the circle graph one copy of the new vertex $v$ between the head and tail of $\alpha$ and the other copy of $v$ between the head and tail of $\beta$ before finally connecting the two copies of $v$ with a new chord.  In terms of the interlace graph, move IV introduces one new vertex $v$ which is connected to all those existing vertices that it is interlaced with in the chord diagram. The reverse operation is just the loopless splitting of vertex $v$ in $\vec{G}$ that is achieved by removing the chord associated with $v$. From this point of view, it is sometimes more convenient to denote by $I( \vec{\Gamma}_{\vec{G}} ) - v$ the interlace graph $I( \vec{\Gamma}_{\vec{H}} )$.            

If a pair of vertices $i$ and $j$ are interlaced in $\vec{\Gamma}$ then the {\emph{transposition}}  $t_{ij}$ of the pair $(ij)$ gives the eulerian circuit $\vec{\Gamma}_{ij} = t_{ij} \vec{\Gamma}$ defined by replacing the vertex sequence $(... i k_1 ... k_p j ... i l_1 ... l_q j ...)$ with $(... i l_1 ... l_q j ... i k_1 ... k_p j ...)$ in $\vec{\Gamma}$ (i.e. exchanging the order of the two sequences of vertices $( k_1 ... k_p )$ and $( l_1 ... l_q )$ which both run from $i$ to $j$). Clearly any pair of vertices which are interlaced in an eulerian circuit in $\vec{H}$ will remain so in the corresponding eulerian circuit in $\vec{G}$ resulting from move IV. Moreover, it is straightforward to check that the act of transposing interlaced vertices commutes with move IV on all elements in $\vec{\mathfrak F}_2$. The effect that transposing interlaced vertices has on the associated interlace graph is described by an operation called {\emph{pivoting}} which is defined as follows. Given an undirected simple graph $H$, begin by selecting a pair of distinct vertices $i$ and $j$ (which we will assume are connected by an edge). Next partition all the vertices other than $i$ and $j$ in $H$ into four sets according to how they are connected to $i$ and $j$ by edges. Vertices which are connected to $i$ but not $j$ live in the first set. Vertices which are connected to $j$ but not $i$ live in the second set. Vertices which are connected to both $i$ and $j$ live in the third set. All the other vertices not connected to either $i$ or $j$ live in the fourth set. The next step is to \lq toggle' pairs of vertices amongst the first three sets. That is, for any pair of vertices such that each vertex in the pair lives in a different one of the first three sets, if the pair is connected by an edge then remove the edge while if the pair is not connected by an edge then connect them with one. The resulting undirected simple graph $H_{ij}$ is called the pivot of $H$ about the edge $ij$. It is not difficult to see that the interlace graph $I ( \vec{\Gamma}_{ij} )$ of the transposed eulerian circuit $\vec{\Gamma}_{ij}$ is then, after relabelling $i \leftrightarrow j$, precisely the pivot of the original interlace graph $I ( \vec{\Gamma} )$ about $ij$. Clearly both transposition and pivoting are involutive operations. Even though $\vec{G}$ is connected, $I( \vec{\Gamma} )$ need not be. However, if $I( \vec{\Gamma} )$ is connected then so is its pivot about any edge.         

The operations described in the previous paragraph are useful in order to understand the structure of circuits in a $2$-regular eulerian digraph. For example, a nice result from lemma 4 in \cite{InterlaceABS} is that any two eulerian circuits in a $2$-regular eulerian digraph are related by some number of transpositions. Thus any two interlace graphs that are related by some number of pivotings must encode the same $2$-regular eulerian digraph. The main result in \cite{InterlaceABS} however is the construction of a well-defined map that associates to the interlace graph $I$ of an eulerian circuit in any $2$-regular eulerian digraph $\vec{G}$ a polynomial function $q_I (x)$ in one variable $x$ with integer coefficients. This function is called the {\emph{interlace polynomial}} and is defined uniquely by the simple recursion relations:
$$
q_I (x) = 
\left\{ 
\begin{array}{ll}
q_{I - i}(x) + q_{I_{ij} - j} (x) & {\mbox{if $ij$ is an edge in $I$,}}\\
x^n & {\mbox{if $I$ has $n$ vertices and no edges.}} \\
\end{array}
\right.
$$      
Now let $r_k ( \vec{G} )$ denote the number of inequivalent partitions of $\vec{G}$ into $k$ (arrow-labelled) circuits and define the {\emph{circuit partition polynomial}} as $r_{\vec{G}}  (x) = \sum_k  r_k ( \vec{G} ) x^k$. These two polynomials are related via $r_{\vec{G}} (x) = x\, q_I (x+1)$ and thus the interlace polynomial provides a useful tool for counting circuit partitions of $2$-regular eulerian digraphs. For example, $q_I (1)$ is the number of eulerian circuits on $\vec{G}$ while $q_I (2) = 2^n$ is number of circuit partitions of $\vec{G}$ defined by pairing up incoming and outgoing arrows at each vertex in $\vec{G}$.  Notice that the relation between $r_{\vec{G}}$ and $q_I$ is well-defined because the interlace polynomial does not depend on the choice of eulerian circuit in $\vec{G}$ that is encoded by $I$ (i.e. it is invariant under pivoting).      

\section{Toric geometry from quiver representations}
\label{sec:toricquivers}

\subsection{Quiver representations from digraphs}
\label{sec:quiverrep}

Let us define a {\emph{quiver representation}} to be the complex unitary representation obtained by associating to each vertex $i$ in a directed graph $\vec{G}$ a complex unitary representation $V_i$ of a real lie group ${\mathcal{G}}_i$. An arrow pointing from vertex $i$ to vertex $j$ in $\vec{G}$ denotes (the real form of) the representation ${\bar V}_i \otimes V_j$. Perhaps the most common class of examples encountered in physical applications have $V_i \cong {\bf N}_i$ corresponding to the fundamental representation of ${\mathcal{G}}_i \cong U( N_i )$, wherein an arrow pointing from vertex $i$ to $j$ corresponds to the real form of the bifundamental representation of $U( N_i ) \times U( N_j )$ while a loop based at vertex $i$ corresponds to the adjoint representation of $U( N_i )$.

For simplicity, we will henceforth only consider the special case where ${\mathcal{G}}$ is abelian and let us assume for the time being that $\vec{G}$ is loopless. If $\vec{G}$ contains $n$ vertices and $e$ arrows then ${\mathcal{G}} \cong U(1)^n$ and $V \cong \CC^e$ as a vector space. Labelling vertices $i=1,...,n$ and arrows $a=1,...,e$ in $\vec{G}$ fixes a basis for the quiver representation. With respect to this basis, the action of ${\mathcal{G}}$ on $V$ is defined by $X_a \mapsto \mbox{exp} \left( i \sum_{i=1}^n \theta_i Q_{ia}  \right) X_a$, for all $( e^{i \theta_1} ,..., e^{i \theta_n} ) \in {\mathcal{G}}$ and $X \in V$, in terms of the $n \times e$ array of integer charges $Q_{ia}$ comprising the so-called {\emph{incidence matrix}} of $\vec{G}$.  A given column $a$ of the incidence matrix encodes an arrow in such a way that it contains an entry $+1$ or $-1$ in row $i$ if the arrow points respectively to or from vertex $i$, or a zero entry if the arrow does not touch $i$. Thus each column contains precisely one $+1$ and one $-1$ entry with only zeroes remaining and hence $\sum_{i=1}^n Q_{ia} =0$ for each $a$.  Notice that this representation would act trivially on loops in $\vec{G}$ and thus could not distinguish between loops based at different vertices, hence their temporary exclusion. The ${\mathcal{G}}$-invariant hermitian inner product is just the canonical complex sesquilinear form $\sum_{a=1}^e X_a {\bar Y}_a$ on $\CC^e$ for all $X,Y \in V$. An important remark to make at this point is that if $\vec{G}$ is chosen to be eulerian, and thus each of its vertices is balanced, then the incidence matrix must also obey $\sum_{a=1}^e Q_{ia} =0$ for each $i$.

The condition $\sum_{i=1}^n Q_{ia} =0$ ensures that the quiver representation defined above is not faithful. That is, for a given arrow $a$ which points from vertex $i$ to vertex $j$, $\sum_{i=1}^n \theta_k Q_{ka} = - \theta_i + \theta_j$ and so the kernel ${\mathcal{K}}$ of the quiver representation has dimension equal to the number of disconnected components of $\vec{G}$ that are not connected to each other by any arrows. If $\vec{G}$ is weakly connected then ${\mathcal{K}}$ is isomorphic to the diagonal $U(1)$ subgroup of $U(1)^n$ containing elements of the form $( e^{i \theta} ,..., e^{i \theta} )$. Henceforth we shall assume that this is always the case. Therefore, one can recover from the quiver representation associated with any weakly connected loopless digraph $\vec{G}$ an effective action of the quotient group ${\mathcal{H}} = {\mathcal{G}}/{\mathcal{K}} \cong U(1)^{n-1}$ on $V$.   

\subsection{Affine toric varieties from digraphs}
\label{sec:affinetoric}

Let us now use the data for the quiver representation above to associate with each weakly connected loopless digraph $\vec{G}$ a {\emph{convex rational polyhedral cone}}
\begin{equation}\label{eq:polycone}
\Lambda_{\vec{G}} = {\mbox{Cone}} ( \Psi_{\vec{G}} ) = \left\{ \sum_{a=1}^e \zeta_a \, {\bm \nu}_a  \, \middle | \, \forall \, \zeta_a \in \RR_{\geq 0} \right\} \subset \RR^{e-n+1}~,
\end{equation}
that is generated by a finite set
\begin{equation}\label{eq:polyconegen}
\Psi_{\vec{G}} = \left\{ {\bm \nu}_a \in \ZZ^{e-n+1} \, \middle | \, \sum_{a=1}^e Q_{ia} \, {\bm \nu}_a = {\bm 0} \right\}~,
\end{equation}
of $e$ integral $(e-n+1)$-vectors subject to precisely $n-1$ linearly independent relations dictated by the incidence matrix of $\vec{G}$. For any choice of $\Psi_{\vec{G}}$, each integral vector ${\bm \nu}_a$  is to be associated with arrow $a$ in $\vec{G}$. We take any such $\Psi_{\vec{G}}$ to be the unique {\emph{minimal rational generating set}} formed by intersecting the $1$-dimensional faces, or {\emph{rays}}, of $\Lambda_{\vec{G}}$ with $\ZZ^{e-n+1}$. It is minimal in the sense that any other generating set for $\Lambda_{\vec{G}}$ must contain $\{ m_a \, {\bm \nu}_a \}$ for some $m_a \in \ZZ_{>0}$. In this situation we take the vectors ${\bm \nu}_a$ to be primitive (i.e. for each $a$, there exists no integer $m_a \neq 1$ such that ${\bm \nu}_a =  m_a \, {\bm \rho}_a$ for any ${\bm \rho}_a \in \ZZ^{e-n+1}$). It is important to stress that this is only fixing how a given generating set $\Psi_{\vec{G}}$ is to be contained in the corresponding polyhedral cone $\Lambda_{\vec{G}}$. However, clearly the relations $\sum_{a=1}^e Q_{ia} \, {\bm \nu}_a = {\bm 0}$ alone are insufficient to determine a unique solution for $\Psi_{\vec{G}}$ and indeed different solutions will typically generate different polyhedral cones. As a first step towards removing this ambiguity, it will be convenient to impose a few more caveats. First of all,  whenever possible, let us choose $\Psi_{\vec{G}}$ such that $\Lambda_{\vec{G}}$ is {\emph{strongly convex}}, meaning that $\Lambda_{\vec{G}}$ and $-\Lambda_{\vec{G}}$ intersect only at the origin in $\RR^{e-n+1}$. If $\Lambda_{\vec{G}}$ is not strongly convex then it must contain a line through the origin and it is easily shown using the relations for $\Psi_{\vec{G}}$ that any point $\sum_{a=1}^e \zeta_a \, {\bm \nu}_a$ on such a line must have vanishing coefficients $\zeta_a = 0$ for each arrow $a$ that is contained in some circuit in $\vec{G}$. Consequently, for any choice of $\Psi_{\vec{G}}$, the corresponding polyhedral cone $\Lambda_{\vec{G}}$ is guaranteed to be strongly convex whenever ${\vec{G}}$ is eulerian. Let us also choose $\Psi_{\vec{G}}$ such that $\Lambda_{\vec{G}}$ has maximal dimension $e-n+1$. In general, the span $\langle \Psi_{\vec{G}} \rangle$ of $\Psi_{\vec{G}}$ over the integers forms a sublattice of $\ZZ^{e-n+1}$ and $\ZZ^{e-n+1} /  \langle \Psi_{\vec{G}}  \rangle$ will be isomorphic to a finite abelian group $\Gamma_{\vec{G}}$ that is non-trivial whenever the sublattice is proper. It will often be convenient in later sections to choose $\Psi_{\vec{G}}$ such that $\langle \Psi_{\vec{G}} \rangle$ is isomorphic to the lattice $\ZZ^{e-n+1}$ so that $\Gamma_{\vec{G}}$ is trivial -- this will ensure that the data from $\vec{G}$ alone is sufficient to specify $\Lambda_{\vec{G}}$ uniquely, up to an affine unimodular transformation of the lattice. However, for the moment let us follow the generic construction wherein $\Gamma_{\vec{G}}$ can be non-trivial.     

In the interests of making our exposition relatively self-contained, we will now briefly summarise the canonical way in which, from the different viewpoints of symplectic and algebraic geometry, the strongly convex rational polyhedral cone $\Lambda_{\vec G}$ can be used to encode the toric structure on a particular K\"{a}hler quotient and affine variety containing a conical singularity. A detailed introduction to symplectic toric geometry (for compact manifolds) and toric varieties in algebraic geometry can be found in, for example, \cite{SilvaSympToric} and \cite{CoxToric,FulToric} respectively. See \cite{GIT3} for a good introduction to the relationship between these two types of toric structures from the perspective of geometric invariant theory.  

The effective action of ${\mathcal{H}}$ that was derived from the quiver representation can be used to construct a K\"{a}hler quotient $\CC^e {/\!/} {\mathcal{H}}$. Taking $\omega = -i \sum_{a=1}^e dX_a \wedge d {\bar X}_a$ to be the canonical K\"{a}hler form on $\CC^e$ with respect to the quiver basis then the $n-1$ linearly independent vector fields $\xi_i = i \sum_{a=1}^e Q_{ia} \left( X_a \frac{\partial}{\partial X_a} - {\bar X}_a \frac{\partial}{\partial {\bar X}_a} \right)$ generate an effective Hamiltonian action of ${\mathcal{H}}$ on $\CC^e$ preserving $\omega$. Thus one can solve $\iota_{\xi_i} \omega = d \mu_i$ to obtain the moment maps
\begin{equation}\label{eq:mmap}
\mu_i (X) = \sum_{a=1}^e Q_{ia} | X_a |^2 - t_i~,
\end{equation}
up to a choice of integration constants $t_i \in \RR$ obeying $\sum_{i=1}^n t_i = 0$. Just as the vector fields $\xi_i$ should naturally be thought to comprise an element in the lie algebra $\fh$ of ${\mathcal{H}}$, so too should the image of the moment maps $\mu_i (X)$ be thought to comprise an element in the dual lie algebra $\fh^* \cong \RR^{n-1}$. The K\"{a}hler quotient is defined as the space of solutions of the $n-1$ independent real equations $\mu_i (X) = 0$ modulo the action of ${\mathcal{H}}$ and is a K\"{a}hler manifold of complex dimension $e-n+1$. Moreover, the K\"{a}hler quotient is {\emph{symplectic toric}} in the sense that the canonical action of $U(1)^e$ on $\CC^e$ descends to an effective Hamiltonian action of the maximal torus $U(1)^e/{\mathcal{H}}$ on $\CC^e {/\!/} {\mathcal{H}}$ preserving the K\"{a}hler structure. The choice of integration constants $t_i$ is an essential piece of extra data that is needed to specify the topology of the K\"{a}hler quotient (e.g. it fixes the K\"{a}hler class) -- they are identified with the Fayet-Iliopoulos parameters in the associated gauged linear sigma model construction discussed in Section~\ref{sec:introduction}. 

Let us denote by ${\mathcal{M}}_{\vec{G}}$ the particular K\"{a}hler quotient defined such that all $t_i =0$. Clearly ${\mathcal{M}}_{\vec{G}}$ is conical since it admits a homothety generated by $X \mapsto \gamma X$ which preserves $\sum_{a=1}^e Q_{ia} | X_a |^2 =0$ for any $\gamma \in \RR_{>0}$. The apex of ${\mathcal{M}}_{\vec{G}}$ is at $X = 0$. It is perhaps worth noting that the other possible K\"{a}hler quotients we could have associated with $\vec{G}$ having some or all $t_i \neq 0$ can be obtained as certain resolutions of the conical singularity in ${\mathcal{M}}_{\vec{G}}$ (this is certainly the generic situation though the singularity may be only partially resolved for some particular choices of $t_i \neq 0$). Since ${\mathcal{M}}_{\vec{G}}$ is symplectic toric, the vector fields generating the action of the maximal torus $U(1)^e/{\mathcal{H}}$ define an $(e-n+1)$-tuple of moment maps taking each point in ${\mathcal{M}}_{\vec{G}}$ to an element in the dual lie algebra of $U(1)^e/{\mathcal{H}}$ which is isomorphic to $\RR^{e-n+1}$.  From the results of \cite{FalMor} (see also \cite{Lerman}), one finds that the image of ${\mathcal{M}}_{\vec{G}}$ under these moment maps is precisely the dual cone
\begin{equation}\label{eq:dualpolycone}
\Lambda_{\vec{G}}^\vee = \left\{ {\bm \upsilon} \in \RR^{e-n+1}  \, \middle | \,  \langle {\bm \upsilon} , {\bm \nu}_a \rangle  \geq 0 \, , \,  \forall \, {\bm \nu_a} \in \Psi_{\vec{G}} \right\}~,
\end{equation}
where $\langle -,- \rangle$ denotes the canonical euclidean inner product on $\RR^{e-n+1}$. This is an equivalent way of defining a cone as the intersection of closed half-spaces and it follows from this definition that $( \Lambda_{\vec{G}}^\vee )^\vee = \Lambda_{\vec{G}}$ with elements in $\Psi_{\vec{G}}$ describing the (inward pointing) normal vectors to the facets of $\Lambda_{\vec{G}}^\vee$. The property of being a strongly convex rational polyhedral cone of maximal dimension is preserved by this duality. It is important to emphasise that the image $\Lambda_{\vec{G}}^\vee$ of ${\mathcal{M}}_{\vec{G}}$ here is only defined up to an affine unimodular  transformation preserving the lattice in which the generating set for $\Lambda_{\vec{G}}$ is contained. The apex of $\Lambda_{\vec{G}}^\vee$ corresponds to the image of the apex of ${\mathcal{M}}_{\vec{G}}$ and can always be fixed to sit at the origin in $\RR^{e-n+1}$ by choosing the integration constants for the moment maps such that they all transform homogeneously under the homothetic action of $\RR_{>0}$ on ${\mathcal{M}}_{\vec{G}}$. Conversely, ${\mathcal{M}}_{\vec{G}}$ can be reconstructed as a torus bundle over $\Lambda_{\vec{G}}^\vee$ that is defined in terms of the data from $\Lambda_{\vec{G}}$. The typical fibre over a generic point in $\Lambda_{\vec{G}}^\vee$ is isomorphic to the maximal torus $U(1)^e/{\mathcal{H}}$ but degenerates to particular $(e-n+1-m)$-dimensional tori over faces in $\Lambda_{\vec{G}}^\vee$ with codimension $m$. The degenerations are encoded by the way the faces sit in $\Lambda_{\vec{G}}^\vee$ (e.g. for $m=1$, the integral components of any normal vector ${\bm \nu_a}$ to a facet in $\Lambda_{\vec{G}}^\vee$ defines the weights for a circle action $U(1)_a$ on $U(1)^e/{\mathcal{H}}$ such that the fibre over that facet is isomorphic to the quotient $( U(1)^e/{\mathcal{H}} ) / U(1)_a$). From this point of view, having fixed the apex of $\Lambda_{\vec{G}}^\vee$ to be at the origin in $\RR^{e-n+1}$, the remaining freedom that allows $\Lambda_{\vec{G}}^\vee$ to be modified by a unimodular transformation in $SL(e-n+1,\ZZ )$  may be interpreted as performing the same transformation as an automorphism of the generic $U(1)^e/{\mathcal{H}}$ fibres in ${\mathcal{M}}_{\vec{G}}$. 

There is also a canonical way to construct an {\emph{affine toric variety}} from a strongly convex rational polyhedral cone $\Lambda_{\vec{G}}$. It proceeds by first defining the map 
\begin{align}\label{eq:pimap}
    \Phi : &\; ( \CC^* )^e \rightarrow ( \CC^* )^{e-n+1} \nonumber \\
    &\; x \mapsto \prod_{a=1}^e x_a^{{\bm \nu}_a}~,
\end{align}
which is a group homomorphism between algebraic tori, thought of as abelian groups with respect to componentwise multiplication (i.e. each $x \in ( \CC^* )^e$ corresponds to an $e$-tuple of non-zero complex numbers with component $x_a$ being associated with arrow $a$ in $\vec{G}$). Since $\Lambda_{\vec{G}}$ is strongly convex, $\Phi$ is surjective and its kernel is isomorphic to ${\mathcal{H}}_\CC \times \Gamma_{\vec{G}}$ which acts effectively on $\CC^e$. The action of ${\mathcal{H}}_\CC \cong ( \CC^* )^{n-1}$ here just follows from the obvious complexification of the the effective action of ${\mathcal{H}}$ on $\CC^e$ defined by the quiver representation. That is, by replacing each $e^{i \theta_i} \in U(1)$ in elements of ${\mathcal{G}}$ with $\lambda_i \in \CC^*$ in elements of ${\mathcal{G}}_\CC$ one obtains the action $X_a \mapsto \left( \prod_{i=1}^n \lambda_i^{Q_{ia}}  \right) X_a$, for all $( \lambda_1 ,..., \lambda_n ) \in {\mathcal{G}}_\CC$ and $X \in V$. Thus, from this construction one is simply recovering the charges $Q_{ia}$ in the incidence matrix for the quiver representation as the coefficients in the linear relations $\sum_{a=1}^e Q_{ia} \, {\bm \nu}_a = {\bm 0}$ amongst the integral vectors in $\Psi_{\vec{G}}$. When $\langle \Psi_{\vec{G}} \rangle$ forms a proper sublattice of $\ZZ^{e-n+1}$, the non-trivial finite abelian group $\Gamma_{\vec{G}} \cong \ZZ^{e-n+1} /  \langle \Psi_{\vec{G}}  \rangle$ is also contained in the kernel of $\Phi$ with the canonical action on $\CC^e$.

The affine toric variety is then defined by considering the holomorphic quotient of $\CC^e$ by ${\mathcal{H}}_\CC \times \Gamma_{\vec{G}}$. However, naively quotienting in this way by a non-compact lie group typically gives rise to rather unpleasant spaces that are not even Hausdorff. The problem is due to the existence of certain orbits  which degenerate under the action of ${\mathcal{H}}_\CC$. However, for toric varieties, one can remedy this problem in a systematic way using the data encoded by $\Lambda_{\vec{G}}$ to remove all the problematic orbits from the quotient. The prescription involves first selecting all the different subsets of $\Psi_{\vec{G}}$ consisting of integral vectors that do not generate a cone in $\Lambda_{\vec{G}}$. Each such subset in $\Psi_{\vec{G}}$ defines an algebraic subset in $\CC^e$ such that, for each ${\bm \nu}_a$ contained in the former, one sets the associated component $X_a =0$ for all $X \in \CC^e$ in the latter. Denoting by $Z_{\Lambda_{\vec{G}}}$ the union of all these algebraic sets then allows one to write the affine toric variety as the holomorphic quotient 
\begin{equation}\label{eq:holquot}
{\mathcal{M}}_{\vec{G}} = \frac{\CC^e \,  \backslash \, Z_{\Lambda_{\vec{G}}}}{{\mathcal{H}}_\CC \times \Gamma_{\vec{G}}}~.
\end{equation}
It is toric in the sense that it contains a maximal algebraic torus isomorphic to $(\CC^* )^{e-n+1}$ as a dense open subset and it inherits a natural action of $( \CC^* )^e / {\mathcal{H}}_\CC$ under the quotient. An alternative realisation of this quotient that is often convenient to work with is as the affine scheme of prime ideals of the coordinate ring of ${\mathcal{M}}_{\vec{G}}$ itself. This coordinate ring is generated by ${\mathcal{H}}_\CC$-invariant monomials of the form $\prod_{a=1}^e X_a^{\langle {\bm \upsilon} , {\bm \nu}_a \rangle}$ for all ${\bm \upsilon} \in \ZZ^{e-n+1} $ such that $\langle {\bm \upsilon} , {\bm \nu}_a \rangle \geq 0$ for all $a=1,...,e$, where $\langle -,- \rangle$ denotes the dual pairing for $\ZZ^{e-n+1}$. Whence, ${\bm \upsilon}$ is necessarily an element in the semigroup $\Lambda_{\vec{G}}^\vee \cap \ZZ^{e-n+1}$ which, from Gordan's lemma, is finitely generated as a consequence of $\Lambda_{\vec{G}}$ being convex rational. Addition in the semigroup $\Lambda_{\vec{G}}^\vee \cap \ZZ^{e-n+1}$ becomes multiplication in the coordinate ring and linear relations amongst the generators of $\Lambda_{\vec{G}}^\vee \cap \ZZ^{e-n+1}$ become monomial relations amongst elements in the coordinate ring.
  
Let us conclude by briefly justifying why ${\mathcal{M}}_{\vec{G}}$ has been used to denote both the conical geometries found in terms of K\"{a}hler and holomorphic quotients constructed from $\Lambda_{\vec{G}}$ above. It is first worth remarking that the reconstruction of ${\mathcal{M}}_{\vec{G}}$ as a K\"{a}hler quotient in terms of a torus bundle over $\Lambda_{\vec{G}}^\vee$ requires some additional data unless ${\mathcal{M}}_{\vec{G}}$ is non-singular everywhere except at its apex. In order for this condition to be met in fact requires that $\Lambda_{\vec{G}}^\vee$ be \lq good' in the sense of Lerman's definition 2.17 in \cite{Lerman} and this necessitates the finite group $\Gamma_{\vec{G}}$ defined in the holomorphic quotient construction being trivial. Assuming for simplicity that this is the case then, at a generic point $X \in \CC^e$ where all $X_a \neq 0$, the equivalence between the two constructions can be seen as follows (this equivalence was first noted in \cite{WittenGLSM} and later generalised in \cite{LutyTaylor} in the context of gauged linear sigma models). The first step is to relate each $e^{i \theta_i}$ in ${\mathcal{G}}$ for the K\"{a}hler quotient with each $\lambda_i$ in ${\mathcal{G}}_\CC$ for the holomorphic quotient such that $\lambda_i = r_i e^{i \theta_i}$ for some $r_i \in \RR_{>0}$. One then finds that the moment map equation $\sum_{a=1}^e Q_{ia} | X_a |^2 =0$, which is invariant under ${\mathcal{G}}$ but not ${\mathcal{G}}_\CC$, can be obtained identically by fixing some particular values for the radial parameters $r_i$ in terms of $X_a$.  In this way one can match up all the coset representatives in the two constructions. 

\subsection{Affine toric Calabi-Yau varieties from eulerian digraphs}
\label{sec:affinetoricCY}

Recall from Section~\ref{sec:quiverrep} that a consequence of $\vec{G}$ being eulerian is that the charges of the incidence matrix must obey $\sum_{a=1}^e Q_{ia} =0$ for each vertex $i$. Conversely, this condition implies that any weakly connected digraph $\vec{G}$ is eulerian. A crucial fact is that this is also precisely the necessary condition for triviality of the canonical bundle on ${\mathcal{M}}_{\vec{G}}$ understood as an affine toric variety. In particular, the first Chern class of ${\mathcal{M}}_{\vec{G}}$ vanishes only if $\sum_{a=1}^e Q_{ia} =0$. Thus one can associate to every loopless eulerian digraph $\vec{G}$ an affine toric Calabi-Yau variety ${\mathcal{M}}_{\vec{G}}$.

A further useful implication of this condition is that it ensures the integral vectors ${\bm \nu}_a$ generating $\Lambda_{\vec{G}}$ must all end on points in a sublattice contained within some particular $\RR^{e-n} \subset \RR^{e-n+1}$ called the {\emph{characteristic hyperplane}}. This fact can be understood in terms of the map $\Phi$ in {\eqref{eq:pimap}}. Since each of the $e-n+1$ components of $\Phi (x)$ is invariant under ${\mathcal{H}}_\CC \times \Gamma_{\vec{G}}$ then, for every element ${\bm \omega} \in \ZZ^{e-n+1}$, one can define an ${\mathcal{H}}_\CC \times \Gamma_{\vec{G}}$-invariant character of the algebraic torus $( \CC^* )^e$ mapping to $\prod_{a=1}^e x_a^{\langle {\bm \omega} , {\bm \nu}_a \rangle}$ from each $x \in ( \CC^* )^e$. Indeed all such characters of $( \CC^* )^e$ follow in this way and the corresponding set of ${\mathcal{H}}_\CC \times \Gamma_{\vec{G}}$-invariant functions they define can be thought of locally as sections of the trivial bundle on ${\mathcal{M}}_{\vec{G}}$. On the other hand, for any $X \in \CC^e$ with all $X_a \neq 0$, $\prod_{a=1}^e X_a$ is ${\mathcal{H}}_\CC$-invariant due to the condition $\sum_{a=1}^e Q_{ia} =0$ and can be thought of locally as the coefficient of a section of the anti-canonical bundle on ${\mathcal{M}}_{\vec{G}}$ whose triviality is precisely the Calabi-Yau condition. Thus, there must exist some particular ${\bm \eta} \in \Lambda_{\vec{G}}^\vee \cap \ZZ^{e-n+1}$ such that $\langle {\bm \eta} , {\bm \nu}_a \rangle =1$ for all $a=1,...,e$ which can be thought of as the normal vector defining the characteristic hyperplane. From the point of view of symplectic geometry, thinking of ${\mathcal{M}}_{\vec{G}}$ as a Calabi-Yau cone over a Sasaki-Einstein manifold, ${\bm \eta}$ also has a natural interpretation; it corresponds to a particular element in the lie algebra of the maximal torus $U(1)^e / {\mathcal{H}}$ that is associated with the canonical {\emph{Reeb}} Killing vector field on the Sasaki-Einstein manifold whose real metric cone is ${\mathcal{M}}_{\vec{G}}$. The vector ${\bm \eta}$ being integral here is non-trivial and implies that the associated Reeb vector generates a locally free circle action on the Sasaki-Einstein geometry, restricting it to be of quasi-regular type.  

For any eulerian digraph $\vec{G}$, intersecting the strongly convex rational polyhedral cone $\Lambda_{\vec{G}}$ with the characteristic hyperplane associated with the integral vector ${\bm \eta}$ defines a {\emph{convex rational polytope}} as the convex hull of the endpoints of the integral vectors in $\Psi_{\vec{G}}$. It will be convenient in the forthcoming analysis to now fix a basis such that ${\bm \eta} = ( {\bm 0} , 1)$ with ${\bm 0} \in \RR^{e-n} \subset \RR^{e-n+1}$. With respect to this basis, the integral vectors in $\Psi_{\vec{G}}$ can be written ${\bm \nu}_a = ( {\bm v}_a , 1 )$ where each ${\bm v}_a \in \ZZ^{e-n} \subset \ZZ^{e-n+1}$. Notice that the linear relations $\sum_{a=1}^e Q_{ia} \, {\bm \nu}_a = {\bm 0}$ and $\sum_{a=1}^e Q_{ia} \, {\bm v}_a = {\bm 0}$ are equivalent due to the condition $\sum_{a=1}^e Q_{ia} =0$. In terms of these integral vectors, we take the more convenient form of the convex rational polytope as the convex hull
\begin{equation}\label{eq:polytope}
\Delta_{\vec{G}} = {\mbox{Conv}} ( \psi_{\vec{G}} ) = \left\{ \sum_{a=1}^e \zeta_a \, {\bm v}_a  \, \middle | \, \forall \, \zeta_a \in \RR_{\geq 0} \, , \, \sum_{a=1}^e \zeta_a = 1 \right\} \subset \RR^{e-n}~,
\end{equation}
of a finite set
\begin{equation}\label{eq:polytopegen}
\psi_{\vec{G}} = \left\{ {\bm v}_a \in \ZZ^{e-n} \, \middle | \, \sum_{a=1}^e Q_{ia} \, {\bm v}_a = {\bm 0} \right\}~.
\end{equation}
This choice of basis leaves unfixed the $SL(e-n, \ZZ )$ subgroup of $SL(e-n+1, \ZZ )$ that is contained in the isotropy group of ${\bm \eta}$ and redefining $\Delta_{\vec{G}}$ by any unimodular transformation in this subgroup will therefore reconstruct the same affine toric Calabi-Yau cone ${\mathcal{M}}_{\vec{G}}$. With respect to this basis, $\langle \Psi_{\vec{G}} \rangle \cong \ZZ^{e-n+1}$ follows from any choice of $\psi_{\vec{G}}$ containing ${\bm 0} \in \RR^{e-n} \subset \RR^{e-n+1}$ and with $\langle \psi_{\vec{G}} \rangle \cong \ZZ^{e-n} \subset \ZZ^{e-n+1}$. 

Having now established the necessary vocabulary, let us move on to consider what effect the four moves which generate all the eulerian digraphs have on the polytopes above which encode the associated toric Calabi-Yau geometries. 

\subsection{Generating toric Calabi-Yau geometries}
\label{sec:gentoricCY}

For any eulerian digraph $\vec{G}$ in $\vec{\mathfrak G}^{[t]}$, the associated affine toric Calabi-Yau cone ${\mathcal{M}}_{\vec{G}}$ has complex dimension $t+1$ and is encoded by the convex rational polytope $\Delta_{\vec{G}} \subset \RR^t$. Therefore, moves I and IV will both increase by one the dimension of the space in which the polytope lives while moves II and III will both preserve this dimension. We will first consider the detailed effect of each of these four moves in order before concluding with a few examples. 

\subsubsection{Move I}
\label{sec:moveI}

As already noted, adding a loop to any vertex in a digraph $\vec{G} \in \vec{\mathfrak G}^{[t]}$ will modify the quiver representation only by adding one extra dimension to $\CC^e$ which is invariant under the action of ${\mathcal{H}}$. The effect of move I on ${\mathcal{M}}_{\vec{G}}$ will therefore retain no information about which vertex in $\vec{G}$ the loop was added to. In terms of the polytope $\Delta_{\vec{G}}$, the effect of move I involves first supplanting the generating set $\psi_{\vec{G}}$ with a set of vectors in $\ZZ^{t+1}$ satisfying the same relations as they did in $\ZZ^t$. To this new set, one further unconstrained vector in $\ZZ^{t+1}$ is then added, corresponding to the loop that was added to $\vec{G}$. One can use an $SL(t+1, \ZZ )$ transformation to take the integral vector associated with the added loop to be $( {\bm 0} , p)$, for some $p \in \ZZ^*$ where ${\bm 0} \in \ZZ^t \subset \ZZ^{t+1}$, with all the other integral vectors taking the form $( {\bm v}_a , 0 )$ such that they live only in the $\ZZ^t \subset \ZZ^{t+1}$ subspace. The value of the non-zero integer $p$ is not specified here and different values give rise to different polytopes that are not related by a unimodular transformation of the lattice (although the sign of $p$ can be taken positive, without loss of generality). Defining the effect of move I unambiguously for a general choice of generating set $\psi_{\vec{G}}$ would therefore require specifying $p$ as extra data. Thus it is more convenient to work within the class of generating sets $\psi_{\vec{G}}$ defined such that they contain ${\bm 0} \in \RR^{t}$ and have $\langle \psi_{\vec{G}} \rangle \cong \ZZ^t$ so that $\Gamma_{\vec{G}}$ is trivial. An unambiguous definition of move I applied within this class is therefore to always take $p=1$. The new polytope after move I can therefore be taken as $\mbox{Conv} \left( ( {\bm v}_a , 0 ) ,( {\bm 0} , 1) \right)$ which will be denoted by $\Pi \left( \Delta_{\vec{G}} \right)$ since it corresponds to what is referred to as the {\emph{pyramid}} over $\Delta_{\vec{G}}$ in the lattice polytope literature. Thus move I simply maps  ${\mathcal{M}}_{\vec{G}} \rightarrow {\mathcal{M}}_{\vec{G}} \times \CC$ (it would map to a cyclic quotient $\left( {\mathcal{M}}_{\vec{G}} \times \CC \right) / \ZZ_{p}$ for any $p>1$).  

\subsubsection{Move II}
\label{sec:moveII}

Consider the subdivision that is depicted in Figure~\ref{fig2} of an arrow $a$ in a digraph $\vec{G} \in \vec{\mathfrak G}^{[t]}$. The effect this has on $\Delta_{\vec{G}}$ is to first replace the integral vector ${\bm v}_a \in \psi_{\vec{G}}$ with a pair of vectors ${\bm v}_b , {\bm v}_c \in \ZZ^t$. The relations between the integral vectors in $\psi_{\vec{G}}$ are only modified for the vertices $v$, $x$ and $w$ that involve the new arrows $b$ and $c$. The relations for vertices $v$ and $w$ are just as they were before but with ${\bm v}_a$ replaced respectively by ${\bm v}_b$ and ${\bm v}_c$. The relation for vertex $x$ then sets ${\bm v}_b = {\bm v}_c$. Therefore the generating set is also $\psi_{\vec{G}}$ after the subdivision. Hence move II does not effect ${\mathcal{M}}_{\vec{G}}$. 

\subsubsection{Move III}
\label{sec:moveIII}

Consider now the contraction shown in Figure~\ref{fig2} of an undirected simple arrow $a$ in a digraph $\vec{G} \in \vec{\mathfrak G}^{[t]}$ that produces the digraph $\vec{G} /a \in \vec{\mathfrak G}^{[t]}$. The effect this has on $\Delta_{\vec{G}}$ is simply to delete one of its vertices corresponding to the removal of ${\bm v}_a$ from the generating set $\psi_{\vec{G}}$ (i.e. $\Delta_{\vec{G} /a} = \mbox{Conv} \left( \psi_{\vec{G}} \, \backslash \,  {\bm v}_a  \right)$). Its removal is of course consistent with the relations between the elements in $\psi_{\vec{G}}$ since one can combine the two relations for the vertices $v$ and $w$ (to which $a$ is attached) into the single relation $\sum_{b \neq a} Q_{vb} \, {\bm v}_b = - \sum_{b \neq a} Q_{wb} \, {\bm v}_b \,  ( = {\bm v}_a )$ for the identified vertex $v=w$. That is, one defines the charges $Q_{(v=w) \, b} = Q_{vb} + Q_{wb}$ for any arrow $b$ in $\vec{G} /a$.  Since $a$ is undirected simple, any arrow $b$ in $\vec{G} /a$ could only have been connected to either $v$ or $w$ or neither in $\vec{G}$ and so $Q_{(v=w) \, b} $ is still either $\pm 1$ or $0$ and defines correctly the corresponding row in the incidence matrix for $\vec{G} /a$. Let us denote by $\CC_a$ the axis in $\CC^e$ associated with arrow $a$ and by $U(1)_{vw}$ the subgroup of ${\mathcal{G}}$ consisting of elements containing the identity for every $U(1)$ factor in ${\mathcal{G}}$ except for the pair associated with vertices $v$ and $w$ whose angles sum to zero. In terms of the quiver representation, move III therefore replaces $\CC^e$ with $\CC^e \backslash \CC_a^*$ and ${\mathcal{G}}$ with ${\mathcal{G}} / U(1)_{vw}$. The subgroup $U(1)_{vw}$ intersects the kernel of the quiver representation only at the identity (since $\vec{G}$ is weakly connected) and so the effective action of ${\mathcal{H}}$ for $\vec{G}$ is replaced with the effective action of ${\mathcal{H}} / U(1)_{vw}$ for $\vec{G} /a$. Furthermore,  since $\psi_{\vec{G}/a} = \psi_{\vec{G}} \, \backslash \,  {\bm v}_a$ for the polytope $\Delta_{\vec{G} /a}$ then clearly $\Psi_{\vec{G}/a} = \Psi_{\vec{G}} \, \backslash \,  {\bm \nu}_a$ for the polyhedral cone $\Lambda_{\vec{G} /a}$ which defines the finite abelian group $\Gamma_{\vec{G} /a} = \ZZ^{t+1} / \langle \Psi_{\vec{G} /a} \rangle$. Thus one arrives at the effect of move III mapping ${\mathcal{M}}_{\vec{G}}$ to
\begin{equation}\label{eq:holquotmoveIII}
{\mathcal{M}}_{\vec{G} /a} = \frac{\left( \CC^e \, \backslash \, \CC_a^* \right) \backslash \, Z_{\Lambda_{\vec{G} /a}}}{\left( {\mathcal{H}}_\CC / \CC_{vw}^* \right) \times \Gamma_{\vec{G} /a}}~.
\end{equation}

This move has a natural physical interpretation in terms of \lq Higgsing' the corresponding superconformal field theory whose gauge-inequivalent D-term vacua are described by ${\mathcal{M}}_{\vec{G}}$. This proceeds by expanding around a constant non-zero vacuum expectation value for the matter field $X_a$ associated with arrow $a$ which breaks the $U(1)_{vw}$ subgroup of the gauge symmetry. A new superconformal field theory is recovered at energy scales much lower than that set by the vacuum expectation value for $X_a$ and the Higgs branch of gauge-inequivalent D-term vacua in this new theory is precisely ${\mathcal{M}}_{\vec{G} /a}$.

\subsubsection{Move IV}
\label{sec:moveIV}

Finally, let us consider the simple immersion shown in Figure~\ref{fig8} of a digraph $\vec{H}  \in \vec{\mathfrak F}_2^{[t]}$ into another digraph $\vec{G}  \in \vec{\mathfrak F}_2^{[t+1]}$. This move defines a map from the polytope $\Delta_{\vec{H}} \subset \RR^t$ to another polytope $\Delta_{\vec{G}} \subset \RR^{t+1}$ in the following way. First notice that identifying all the arrows $\gamma$ which $\vec{H}$ and $\vec{G}$ have in common implies there must exist a bijection between the elements of $\psi_{\vec{H}} \backslash \{ {\bm v}_\alpha , {\bm v}_\beta \}$ in $\ZZ^t$ and those of $\psi_{\vec{G}} \backslash \{ {\bm v}_a , {\bm v}_b , {\bm v}_c , {\bm v}_d \}$ in $\ZZ^{t+1}$ such that $\ZZ^t$ is embedded as a sublattice in $\ZZ^{t+1}$. It is convenient to use an $SL(t+1, \ZZ )$ transformation to fix a basis such that the bijection is defined by writing each integral vector in $\psi_{\vec{G}} \backslash \{ {\bm v}_a , {\bm v}_b , {\bm v}_c , {\bm v}_d \}$ that is associated with an arrow $\gamma$ as $( {\bm v}_\gamma , w_\gamma )$, where ${\bm v}_\gamma$ is the corresponding integral vector in  $\psi_{\vec{H}} \backslash \{ {\bm v}_\alpha , {\bm v}_\beta \}$ and $w_\gamma$ is an integer specifying the coordinate in the complementary direction to $\ZZ^t \subset \ZZ^{t+1}$. With respect to this basis, the remaining integral vectors in $\psi_{\vec{G}}$ associated with the four arrows $a$, $b$, $c$ and $d$ respectively become $( {\bm v}_\alpha , w_a )$, $( {\bm v}_\beta , w_b )$, $( {\bm v}_\alpha , w_c )$ and $( {\bm v}_\beta , w_d )$ for some integers $w_a$, $w_b$, $w_c$ and $w_d$, such that one recovers the polytope $\Delta_{\vec{H}}$ from $\Delta_{\vec{G}}$ by projecting onto the corresponding $\RR^t \subset \RR^{t+1}$ subspace. This projection just corresponds to the loopless splitting of vertex $v$ in $\vec{G}$ such that the respective heads of  $a$ and $b$ are reconnected to the tails of $c$ and $d$ to recover the arrows $\alpha$ and $\beta$ in $\vec{H}$. Of course, there are many possible choices of integers $\{ w_a , w_b , w_c , w_ d , w_\gamma \}$ for the integral vectors in $\psi_{\vec{G}}$. It is therefore more convenient to work within the class of generating sets $\psi_{\vec{H}}$ defined such that they contain ${\bm 0} \in \RR^{t}$ and have $\langle \psi_{\vec{H}} \rangle \cong \ZZ^t$ so that $\Gamma_{\vec{H}}$ is trivial. By taking each integer in $\{ w_a , w_b , w_c , w_ d , w_\gamma \}$ to be either $0$ or $1$, it will be seen that one is guaranteed to remain within this class of generating sets after applying move IV (i.e. $\psi_{\vec{G}}$ contains $( {\bm 0} , 0) \in \RR^{t+1}$ and $\langle \psi_{\vec{G}} \rangle \cong \ZZ^{t+1}$ so that $\Gamma_{\vec{G}}$ is trivial). We exclude the possibility that these integers all equal $0$ or all equal $1$ because $\Delta_{\vec{G}}$ would then degenerate to a $t$-dimensional polytope in $\RR^{t+1}$ that is just related to $\Delta_{\vec{H}}$ by a lattice translation. Moreover, so that the integral vectors associated with each pair of arrows $(a,c)$ and $(b,d)$ are not identical in $\ZZ^{t+1}$, let us demand that $w_ a + w_c = 1 = w_b + w_d$. 

To examine the relations which must be obeyed by these integral vectors in $\psi_{\vec{G}}$, it will be convenient to introduce the label $A \in \{ a,b,c,d, \gamma \}$ for the arrows in $\vec{G}$. It is clear that the relations for all the components ${\bm v}_\alpha$, ${\bm v}_\beta$, ${\bm v}_\gamma$  in the $\ZZ^t \subset \ZZ^{t+1}$ sublattice for these integral vectors are precisely equivalent to those defining $\psi_{\vec{H}}$ and so are satisfied identically in $\psi_{\vec{G}}$. The only non-trivial relations that must be solved are for the binary integers $w_A \in \{ 0 , 1 \}$ and it will be helpful to think of the value $0$ or $1$ to correspond to the assignment of a colour, say, white or black to the arrow $A$. Since $\vec{G} \in \vec{\mathfrak F}_2^{[t+1]}$, at each vertex $x$ in $\vec{G}$, all these relations must take the form $w_{A_1} + w_{A_2} = w_{A_3} + w_{A_4}$ where the arrows $A_1$, $A_2$ both point to/from $x$ while arrows $A_3$, $A_4$ both point from/to $x$. Consequently, at each vertex, solutions must always have equal numbers of incoming and outgoing arrows with the same colour. Equivalently, for every incoming arrow with a given colour at any vertex in $\vec{G}$, there must be another outgoing arrow from that vertex with the same colour. This implies that arrows of a given colour must traverse circuits in $\vec{G}$ and so the general solution is specified by taking any circuit decomposition for $\vec{G}$ (which must exist since it is eulerian) and colouring each circuit either white or black. Of course, in addition, we must pay heed to the caveats that were assumed at the end of the previous paragraph. Firstly this means that we must not choose the same colour for all the circuits in the decomposition (so the decomposition must contain at least two circuits). Secondly, given that $w_ a + w_c = 1 = w_b + w_d$, the relation $w_a + w_b = w_c + w_d$ for the new vertex $v$ in $\vec{G}$ implies that  both $a$ and $d$ must have the same colour whilst $b$ and $c$ must both have the opposite colour (for concreteness, let us say that $w_a = 0 = w_d$  and $w_b = 1 = w_c$ which represents no loss of generality given that we can always relabel $0 \leftrightarrow 1$ by the binary involution $w_A \mapsto 1 - w_A$). A canonical solution for all the binary integers $w_\gamma$ associated with the arrows $\gamma$ appearing in the remaining relations can then be found by utilising the results described in Section~\ref{sec:encoding2reg} for $2$-regular eulerian digraphs. The solution is obtained by first choosing any eulerian circuit in $\vec{H}$, which must involve some sequence of arrows of the form $( ... \alpha ... \beta ...)$. Recall that the operation of simple immersion can be understood by simply replacing this sequence with $( ... a v c ... b v d ...)$ to form an eulerian circuit in $\vec{G}$. In terms of the chord diagrams for the circle graphs associated with these eulerian circuits, the simple immersion of $\vec{H}$ in $\vec{G}$ therefore just involves inserting the new vertex $v$ at two points bisecting the arrows $\alpha$ and $\beta$ on the circumference and then connecting these two points with a new chord. Rotating the circle graph such that this new chord for $v$ is aligned vertically then the solution follows by colouring respectively white/black all the arrows around the circumference of the circle to the left/right of the chord. The choice of colour scheme is fixed by our having taken $a$ and $d$ to be white and $b$ and $c$ to be black. Furthermore, any vertex that is interlaced with $v$ in the eulerian circuit in $\vec{G}$ must have both pairs of incoming and outgoing arrows with opposite colours. Any vertex that is not interlaced with $v$ must have all four connecting arrows the same colour (the colour being white/black according to whether they lie to the left/right of $v$ in the circle graph).          

By partitioning the set of arrows $\{ \gamma \}$ in $\vec{G}$ into two subsets $\{ \gamma^\circ \, | \, w_{\gamma^\circ} = 0 \}$ and $\{ \gamma^\bullet \, | \, w_{\gamma^\bullet} = 1 \}$ according to the colour assignments defined by the eulerian circuit above,  one can express the polytope $\Delta_{\vec{G}} = \mbox{Conv} \left( ( {\bm v}_\alpha , 0 ), ( {\bm v}_\beta , 1 ), ( {\bm v}_\alpha , 1 ) , ( {\bm v}_\beta , 0 ) , ( {\bm v}_{\gamma^\circ} , 0 ) , ( {\bm v}_{\gamma^\bullet} , 1 ) \right) \subset \RR^{t+1}$. The convex hull of the integral vectors associated with arrows $a$, $b$, $c$ and $d$ is therefore always a rectangle in $\Delta_{\vec{G}}$ which projects to the interval $\mbox{Conv} \left(  {\bm v}_\alpha , {\bm v}_\beta \right)$ in $\Delta_{\vec{H}}$. Polytopes of this kind can be understood as a subclass of a well-studied class of lattice polytopes that is defined as follows. Given a set of $s+1$ lattice polytopes $\{ \Delta_0 , \Delta_1 ,..., \Delta_s \}$ in $\RR^t$, take $\{ {\bf e}_0 , {\bf e}_1 ,..., {\bf e}_s \}$ to define the vertices of the unit simplex in $\RR^s$ such that ${\bf e}_0$ corresponds to the origin in $\RR^s$. From this data, one can define the {\emph{Cayley polytope}} $\Delta_0 * \Delta_1 *...* \Delta_s = \mbox{Conv} \left( \Delta_0 \times \{ {\bf e}_0 \} , \Delta_1 \times \{ {\bf e}_1 \} ,..., \Delta_s \times \{ {\bf e}_s \}  \right)$ as a lattice polytope in $\RR^t \oplus \RR^s$. For example, the pyramid $\Pi ( \Delta )$ over a lattice polytope $\Delta \subset \RR^t$ can be thought of as the Cayley polytope $\Delta * \{ {\bm 0} \}$, where ${\bm 0} \in \RR^t$ denotes the origin. Thus, by defining the pair of polytopes $\Delta_{\vec{G}}^\circ = \mbox{Conv} \left( {\bm v}_\alpha , {\bm v}_\beta , {\bm v}_{\gamma^\circ} \right)$ and $\Delta_{\vec{G}}^\bullet = \mbox{Conv} \left( {\bm v}_\alpha , {\bm v}_\beta , {\bm v}_{\gamma^\bullet} \right)$ in $\RR^t$, one can express $\Delta_{\vec{G}} = \Delta_{\vec{G}}^\circ * \Delta_{\vec{G}}^\bullet$ as a Cayley polytope in $\RR^{t+1}$. Notice that the generating set $\psi_{\vec{G}}$ resulting from move IV has $\langle \psi_{\vec{G}} \rangle \cong \ZZ^{t+1}$, having taken $\langle \psi_{\vec{H}} \rangle \cong \ZZ^{t}$ with the extra lattice direction spanned by the difference between the integral vectors associated with either arrows $a$ and $c$ or arrows $b$ and $d$. Since $\psi_{\vec{H}}$ was taken to contain ${\bm 0} \in \RR^{t}$ then, assuming the arrow associated with this integral vector is coloured white following move IV, $\psi_{\vec{G}}$ must contain $( {\bm 0} , 0) \in \RR^{t+1}$ (one can always use the binary involution relabelling $0 \leftrightarrow 1$ to define the white/black colour assignments such that this is the case).

Notice that the initial polytope $\Delta_{\vec{H}} = \mbox{Conv} \left( {\bm v}_\alpha , {\bm v}_\beta , {\bm v}_\gamma \right)$ will therefore generally not be contained in $\Delta_{\vec{G}}$. In fact, the only way to have $\Delta_{\vec{H}}$ contained as a facet in $\Delta_{\vec{G}}$ here would be if it could be identified with either $\Delta_{\vec{G}}^\circ$ or  $\Delta_{\vec{G}}^\bullet$ which is only possible if all the arrows $\gamma$ could be chosen to have the same colour. Equivalently, this means that the chord associated with the new vertex $v$ introduced by simple immersion must have only arrows $a$ and $d$ to its left or only arrows $b$ and $c$ to its right in the circle graph. In either case, the polytope is given by $\Delta_{\vec{G}} = \Delta_{\vec{H}} * \mbox{Conv} \left(  {\bm v}_\alpha , {\bm v}_\beta \right)$. Moreover, in this special situation, as an alternative to performing the loopless splitting of $v$, one can equivalently recover the polytope $\Delta_{\vec{H}}$ by using a combination of moves III and I. This works by first contracting one of the pair of arrows in $\vec{G}$ that have the opposite colour to all the arrows $\gamma$. This corresponds to move III although it is worth noting that both arrows in the aforementioned pair necessarily connect the vertex $v$ and one other vertex in $\vec{G}$. Hence contracting either one of these two arrows must turn the other into a loop based at the vertex which $v$ is identified with in the contraction. The polytope associated with this intermediate eulerian digraph corresponds to the pyramid $\Pi ( \Delta_{\vec{H}} )$. The final step is the removal of this loop and the resulting polytope $\Delta_{\vec{H}}$ just comes from collapsing the pyramid.   

Before moving on, it is worth making a few remarks about the consistency of this prescription. As mentioned in Section~\ref{sec:encoding2reg}, given any pair of vertices $(ij)$ that are interlaced in an eulerian circuit in $\vec{H}$, the transposition $t_{ij}$ commutes with move IV. Since any two eulerian circuits in $\vec{H}$ are related by some number of transpositions of interlaced vertices, this allowed the simple immersion of $\vec{H}$ in $\vec{G}$ to be defined unambiguously, for any choice of eulerian circuit. However, in the prescription above, an additional piece of data is defined by the subsequent white/black colour scheme assigned to arrows in $\vec{G}$ after performing move IV that is used to define the polytope $\Delta_{\vec{G}}$. Let us take this arrow colour assignment to be part of the definition of move IV and let ${\mathrm{IV}}_v$ denote this operation on an eulerian circuit in $\vec{H}$ such that the new vertex $v$ is introduced in $\vec{G}$. Generally the two $t_{ij}$-conjugate operations ${\mathrm{IV}}_v$ and $t_{ij} \, {\mathrm{IV}}_v \, t_{ij}$ on some eulerian circuit in $\vec{H}$ for which vertices $i$ and $j$ are interlaced will give rise to different coloured eulerian circuits in $\vec{G}$. That is, they still describe the same fixed sequence of labelled vertices and arrows but the assignment of white/black colours to the arrows in $\vec{G}$ will generally be different (they are only the same when $v$ is interlaced with neither $i$ nor $j$). However, it is important to stress that, by construction, the different coloured eulerian circuits in $\vec{G}$ resulting from ${\mathrm{IV}}_v$ and $t_{ij} \, {\mathrm{IV}}_v \, t_{ij}$ both generate solutions of the linear relations in \eqref{eq:polytopegen} for $\Delta_{\vec{G}}$. Consequently, choosing different eulerian circuits in $\vec{H}$ will generally result in different polytopes though they will always be related via some unimodular transformation and therefore give rise to the same affine toric Calabi-Yau variety ${\mathcal{M}}_{\vec{G}}$. One is therefore free to choose any eulerian circuit in $\vec{H}$ from which the colouring assignment prescribed above yields a convenient representative polytope $\Delta_{\vec{G}}$ encoding ${\mathcal{M}}_{\vec{G}}$.

\subsubsection{Examples}
\label{sec:examples}

As explained at the end of Section~\ref{sec:split2regvert}, the $2$-regular eulerian digraphs $\vec{O}_p$ depicted in Figure~\ref{fig7} which do not admit a loopless splitting of any of their vertices can be generated by a particular combination of moves II+I+IV. Using this fact together with the results above allows one to construct the associated polytopes $\Delta_{\vec{O}_p} \subset \RR^{2p}$ recursively in the following way. For any $\vec{O}_p$, the first step is to perform move II by subdividing any one of its undirected simple arrows, which will not modify the polytope $\Delta_{\vec{O}_p}$. The next step is to perform move I by adding a loop based at the new vertex, and the associated polytope is given by the pyramid $\Pi( \Delta_{\vec{O}_p} ) \subset \RR^{2p+1}$. The final step requires first a choice of eulerian circuit in this intermediate digraph. However, from Figure~\ref{fig7} it is clear that any eulerian circuit in $\vec{O}_p$ must traverse vertices in the sequence $(12123434...(2p-1)(2p)(2p-1)(2p))$. Since the subdivided arrow must have pointed from a vertex $2i$ to $2i+1$ in this sequence (for some $i \in \{ 1,...,p \}$ and identifying labels modulo $2p$) then the effect of the previous moves II+I here is just to replace $(...(2i)(2i+1)...)$ with $(...(2i)ww(2i+1)...)$, where $w$ is the label for the new vertex. One must then identify the loop based at $w$ with $\alpha$ and the arrow whose tail is attached to $w$ with $\beta$ to perform the relevant simple immersion corresponding to the final move IV. Thus, the relevant vertex and arrow labels must appear in the order $(...(2i)w \alpha w \beta (2i+1)...)$ in the eulerian circuit which gets replaced with $(...(2i)w a v c w b v d (2i+1)...)$ under  move IV, following again the notation in Figure~\ref{fig8}. This means that only arrows $b$ and $c$ are coloured black in the resulting eulerian circuit in $\vec{O}_{p+1}$ which has resulted from applying this combination of moves II+I+IV to $\vec{O}_p$. Therefore the polytope for $\vec{O}_{p+1}$ can be written $\Delta_{\vec{O}_{p+1}} = \Pi ( \Delta_{\vec{O}_p} ) * \mbox{Conv} \left(  {\bm v}_\alpha , {\bm v}_\beta \right) \subset \RR^{2(p+1)}$. It is worth remarking that it is precisely the four arrows $a$, $b$, $c$ and $d$ in this construction that describe the extra block in $\vec{O}_{p+1}$ relative to $\vec{O}_p$ and that $\mbox{Conv} \left(  {\bm v}_a , {\bm v}_b , {\bm v}_c , {\bm v}_d \right)$ lies in a plane $\RR^2 \subset \RR^{2(p+1)}$ which intersects $\Delta_{\vec{O}_p}$ only at the lattice point ${\bm v}_d$. It is convenient to take this point to be the origin in $\RR^{2(p+1)}$ and so $\mbox{Conv} \left(  {\bm v}_a , {\bm v}_b , {\bm v}_c , {\bm v}_d \right)$ can be taken to describe the unit square in the aforementioned plane. Up to a unimodular transformation, the structure of the general polytope $\Delta_{\vec{O}_p}$ can therefore be obtained by iterating this result. That is, by thinking of $\vec{O}_p$ as a chain of $p$ blocks containing 4 arrows each as in Figure~\ref{fig7}, one associates to each block the unit square in a plane $\RR_i^2 \subset \RR^{2p}$ such that  $\bigcap_{i=1}^p \RR_i^2 = {\bm 0} \in \RR^{2p}$ and then $\Delta_{\vec{O}_p}$ is just the convex hull of the corners of all these squares. Clearly the generating set $\psi_{\vec{O}_p}$ contains ${\bm 0} \in \RR^{2p}$ and has $\langle \psi_{\vec{O}_p} \rangle \cong \ZZ^{2p}$ so that $\Gamma_{\vec{O}_p}$ is trivial. 
  
A similar recursive technique can be utilised to describe the structure of the polytopes associated with the second class of $2$-regular eulerian \lq necklace' digraphs shown in Figure~\ref{fig5} which contain no undirected simple arrows. Let us denote by $\vec{A}_t$ the necklace digraph of this type on $t$ vertices. By selecting any pair of oppositely oriented arrows connecting vertices $i$ and $i+1$ in $\vec{A}_t$ (for some $i \in \{ 1,...,t \}$ and identifying labels modulo $t$) then performing move IV on them will produce the necklace digraph $\vec{A}_{t+1}$. If one identifies this pair of arrows with $\alpha$ and $\beta$ in Figure~\ref{fig8} then one can define an eulerian circuit in $\vec{A}_t$ by the sequence $(... (i-1) i \alpha (i+1) \beta i (i-1) ... (i+2)(i+1)(i+2)...)$ (displaying only the labels for arrows $\alpha$ and $\beta$). The chord associated with the vertex $i+1$ in the circle graph for this eulerian circuit is therefore interlaced with the chords for all the other vertices though none of these other chords are interlaced with each other. Performing move IV replaces the sequence with $(... (i-1) i avc (i+1) bvd i (i-1) ... (i+2)(i+1)(i+2)...)$ which defines an eulerian circuit in $\vec{A}_{t+1}$. Again, the chord for vertex $i+1$ is interlaced with all the others, including the new one for vertex $v$, while none of these other chords are interlaced with each other. Thus, only arrows $b$ and $c$ are to be coloured black here and so the associated polytope can be written $\Delta_{\vec{A}_{t+1}} = \Delta_{\vec{A}_t} * \mbox{Conv} \left(  {\bm v}_\alpha , {\bm v}_\beta \right) \subset \RR^{t+1}$ and one can take $\mbox{Conv} \left(  {\bm v}_\alpha , {\bm v}_\beta \right)$ to be the unit interval $[0,1]$. Consequently, one can express $\Delta_{\vec{A}_t} = \underbrace{[0,1] * [0,1] * ... * [0,1]}_{t}  \subset \RR^t$. This is known as a (narrow) {\emph{Lawrence prism}} in the lattice polytope literature and can be equivalently expressed as $\Delta_{\vec{A}_t} = \sigma_{t-1} * \sigma_{t-1}$ in terms of the unit simplex $\sigma_{t-1} \subset \RR^{t-1}$.  The generating set $\psi_{\vec{A}_t}$ therefore contains ${\bm 0} \in \RR^{t}$ and has $\langle \psi_{\vec{A}_t} \rangle \cong \ZZ^{t}$ since the vertices of either copy of $\sigma_{t-1}$ together with the unit interval connecting them form a $\ZZ$-basis. 
  
The polytopes associated with the first class of $2$-regular eulerian \lq necklace' digraphs in Figure~\ref{fig5} can be described in a similar manner. Denoting by $\vec{B}_t$ the necklace digraph of this type on $t$ vertices then performing move IV on any pair of arrows connecting vertices $i$ and $i+1$ in $\vec{B}_t$ (for some $i \in \{ 1,...,t \}$ and again identifying labels modulo $t$) will produce $\vec{B}_{t+1}$. Identifying this pair of arrows with $\alpha$ and $\beta$ in Figure~\ref{fig8} then they must appear in the sequence $(... (i-1) i \alpha (i+1) (i+2) ... (i-1) i \beta (i+1) (i+2) ...)$ in an eulerian circuit in $\vec{B}_t$ (i.e. ignoring arrow labels, there is only one possible eulerian circuit in $\vec{B}_t$ formed by traversing all the vertices in order twice). The interlace graph for this eulerian circuit in $\vec{B}_t$ is therefore the complete graph $K_t$ on $t$ vertices and move IV replaces the sequence with $(... (i-1) i avc (i+1) (i+2) ... (i-1) i bvd (i+1) (i+2) ...)$ defining the corresponding eulerian circuit in $\vec{B}_{t+1}$ whose interlace graph is $K_{t+1}$. Let us now introduce the labels $i$ and $t+i$ for the pairs of arrows connecting vertices $i$ and $i+1$ in $\vec{B}_t$. The linear relations in \eqref{eq:polytopegen} for the polytope $\Delta_{\vec{B}_t}$ say that the integral vectors must obey $\bm{v}_i + \bm{v}_{t+i} = \bm{u}$ for all $i=1,...,t$ in terms of some fixed $\bm{u} \in \ZZ^t$. The corresponding equations for $\Delta_{\vec{B}_{t+1}}$ following move IV say that each pair of arrows $i$ and $t+i$ in $\vec{B}_t$ must have opposite colours (i.e. $w_i + w_{t+i} = 1$ for all $i=1,...,t$). Since any $\vec{B}_t$ can be constructed by simply repeating this operation then clearly one can take the vector $\bm{u}$ above to contain only unit entries with all the components of $\bm{v}_i$ and $\bm{v}_{t+i}$ being either zero or one. Consequently, for each $i$, the pair of vectors $\bm{v}_i$ and $\bm{v}_{t+i}$ can be taken to end on opposite corners of the unit hypercube $[0,1]^t \subset \RR^t$ and the representative polytope $\Delta_{\vec{B}_t}$ obtained from this construction is defined by $\bm{v}_1 = {\bf e}_0$ and $\bm{v}_i = \sum_{j=2}^i {\bf e}_j$ for $i=2,...,t$, where $\{ {\bf e}_0 , {\bf e}_1 ,..., {\bf e}_t \}$ define the vertices of the unit simplex $\sigma_t \subset \RR^t$ with ${\bf e}_0$ corresponding to the origin in $\RR^t$. The generating set $\psi_{\vec{B}_t}$ clearly has $\langle \psi_{\vec{B}_t} \rangle \cong \ZZ^{t}$ with ${\bf e}_1 = \bm{v}_{2t}$ and ${\bf e}_i = \bm{v}_i - \bm{v}_{i-1}$ for $i=2,...,t$. The corresponding toric Calabi-Yau variety ${\mathcal{M}}_{\vec{B}_t}$ can be expressed as the real metric cone over the compact homogeneous Sasaki-Einstein manifold $SU(2)^t / U(1)^{t-1}$. This can be easily seen by considering the intersection of solutions of the moment map equation \eqref{eq:mmap} for $\vec{B}_t$ (i.e. $| X_i |^2 + | X_{t+i} |^2 = r$ for all $i=1,...,t$ in terms of some $r \in \RR_{>0}$) with the round unit-radius $(4t-1)$-sphere canonically embedded in $\CC^{2t}$ via $\sum_{i=1}^t | X_i |^2 + | X_{t+i} |^2 = 1$. The intersection fixes $r=1/t$ and describes the product of $t$ copies of $S^3 \cong SU(2)$ (each with radius $\frac{1}{\sqrt{t}}$). The quotient by $U(1)^{t-1}$ then follows in the K\"{a}hler quotient construction of ${\mathcal{M}}_{\vec{B}_t}$ as precisely the group ${\mathcal{H}}$ which acts effectively on the intersection above, corresponding to the quotient of the maximal torus $U(1)^t \subset SU(2)^t$ by its diagonal $U(1)$ subgroup. For $t=2,3$, these homogeneous Sasaki-Einstein geometries are often denoted by $T^{1,1}$ and $Q^{1,1,1}$ in the physics literature and appear in the well-known supersymmetric $AdS_5 \times T^{1,1}$ and $AdS_4 \times Q^{1,1,1}$ solutions of IIB string theory and M-theory respectively.  
 
\section*{Acknowledgments}

I would like to thank Jos\'{e} Figueroa-O'Farrill, Amihay Hanany and Dario Martelli for some helpful discussions, Charles Boyer for email correspondence and Elena M\'{e}ndez-Escobar for collaboration during the early stages of this work. I am also grateful to the members of the Institute for the Physics and Mathematics of the Universe (IPMU) for their kind hospitality during the conception of this work. This research was supported in part by grant ST/G000514/1 ``String Theory Scotland'' from the UK Science and Technology Facilities Council. 

\bibliographystyle{utphys}
\bibliography{CYQuiversGLSM}

\end{document}